\documentclass[fleqn,usenatbib]{mnras}

\usepackage{newtxtext,newtxmath}

\usepackage[T1]{fontenc}

\DeclareRobustCommand{\VAN}[3]{#2}
\let\VANthebibliography\thebibliography
\def\thebibliography{\DeclareRobustCommand{\VAN}[3]{##3}\VANthebibliography}

\usepackage{graphicx}	
\usepackage{amsmath}	

\usepackage{amssymb}	
\usepackage{threeparttable}
\usepackage{subfigure}
\usepackage{verbatim}
\usepackage{multirow}

\title[Observation of 4U 0115+63 during 2017 outburst]{QPOs and Orbital elements of X-ray binary 4U 0115+63 during the 2017 outburst observed by \textit{Insight}-HXMT}

\author[Y. Ding et al.]{Y. Z. Ding,$^{1,2}$
W. Wang,$^{1,3}$\thanks{E-mail:wangwei2017@whu.edu.cn} P. Zhang,$^{1,3}$ Q.C. Bu,$^{4}$ C. Cai,$^{4}$ X.L.Cao,$^{4}$ C. Zhi,$^{4}$ L. Chen,$^{5}$ T. X. Chen,$^{4}$
\newauthor
Y. B. Chen,$^{6}$ Y. Chen,$^{4}$ Y. P. Chen,$^{4}$ W. W. Cui,$^{4}$ Y. Y. Du,$^{4}$ G. H. Gao,$^{4,7}$ H. Gao,$^{4,7}$ M. Y. Ge,$^{4}$ Y. D. Gu,$^{4}$
\newauthor
J. Guan,$^{4}$ C. C. Guo,$^{4,7}$ D. W. Han,$^{4}$ Y. Huang,$^{4}$ J. Huo,$^{4}$ S. M. Jia,$^{4}$ W. C. Jiang,$^{4}$ J. Jin,$^{4}$ L. D. Kong,$^{4,7}$
\newauthor
B. Li,$^{4}$ C. K. Li,$^{4}$ G. Li,$^{4}$ T. P. Li,$^{4,7}$ W. Li,$^{4}$ X. Li,$^{4}$ X. B. Li,$^{4}$ X. F. Li,$^{4}$ Z. W. Li,$^{4}$ X. H. Liang,$^{4}$ J. Y. Liao,$^{4}$
\newauthor
B. S. Liu,$^{4}$ C. Z. Liu,$^{4}$ H. X. Liu,$^{4,7}$ H. W. Liu,$^{4}$ X. J. Liu,$^{4}$ F. J. Lu,$^{4}$ X. F. Lu,$^{4}$Q. L.,$^{4,7}$ L. T.,$^{4}$ R. C. Ma,$^{4,7}$
\newauthor
X. Ma,$^{4}$ B. Meng,$^{4}$ Y. Nang,$^{4,7}$ J. Y. Nie,$^{4}$ J. L. Qu,$^{4}$ X. Q. Ren,$^{4}$ N. Sai,$^{4,7}$ L. M. Song,$^{4,7}$ X. Y. Song,$^{4}$
\newauthor
L. Sun,$^{4}$ Y. Tan,$^{4}$ L. Tao,$^{4}$ Y. L. Tuo,$^{4,7}$ L. J. Wang,$^{4}$ P. J. Wang,$^{4,7}$ W. S. Wang,$^{9}$Y. S. Wang,$^{4}$ X. Y. Wen,$^{4}$
\newauthor
B. Y. Wu,$^{4,7}$B. B. Wu,$^{4}$ M. Wu,$^{4}$G. C. Xiao,$^{4,7}$ S. Xiao,$^{4,7}$ S. L. Xiong,$^{4}$ Y. P. Xu,$^{4,7}$ R. J. Yang,$^{11}$ S. Yang,$^{4}$
\newauthor
Y. J. Yang,$^{4}$ Q. B. Yi,$^{4,12}$ Q. Q. Yin,$^{4}$ Y. You,$^{4}$ F. Zhang,$^{4}$ H. M. Zhang,$^{8}$ J. Zhang,$^{4}$ P. Zhang,$^{4}$ S. Zhang,$^{4}$
\newauthor
S. N. Zhang,$^{4,8}$W. C. Zhang,$^{4}$ W. Zhang,$^{4,7}$ Y. F. Zhang,$^{4}$ Y. H. Zhang,$^{4,7}$ H. S. Zhao,$^{4}$ X. F. Zhao,$^{4,7}$
\newauthor
S. J. Zheng,$^{4}$ Y. G. Zheng$^{4,11}$ and D. K. Zhou.$^{4,7}$
\\
$^{1}$Department of Physics and Technology, Wuhan University, Wuhan 430072, China\\
$^{2}$Hongyi Honor school, Wuhan University, Wuhan 430072, China\\
$^{3}$WHU-NAOC Joint Center for Astronomy, Wuhan University, Wuhan 430072, China\\
$^{4}$Key Laboratory of Particle Astrophysics, Institute of High Energy Physics, Chinese Academy of Sciences, 19B Yuquan Road, Beijing 100049, China\\
$^{5}$Department of Astronomy, Beijing Normal University, Beijing 100088, China\\
$^{6}$Department of Physics, Tsinghua University, Beijing 100084,  China\\
$^{7}$University of Chinese Academy of Sciences, Chinese Academy of Sciences, Beijing 100049, China\\
$^{8}$Department of Astronomy, Tsinghua University, Beijing 100084, China\\
$^{9}$Computing Division, Institute of High Energy Physics, Chinese Academy of Sciences, 19B Yuquan Road, Beijing 100049, China\\
$^{10}$Key Laboratory of Space Astronomy and Technology, National Astronomical Observatories, Chinese
Academy of Sciences, Beijing 100012, China\\
$^{11}$College of physics Sciences \& Technology, Hebei University, No. 180 Wusi Dong Road, Lian Chi District, Baoding City, Hebei  071002, China\\
$^{12}$School of Physics and Optoelectronics, Xiangtan University, Yuhu District, Xiangtan, Hunan, 411105, China\\
}

\date{Accepted XXX. Received YYY; in original form ZZZ}

\pubyear{2020}

\begin{document}
\label{firstpage}
\pagerange{\pageref{firstpage}--\pageref{lastpage}}
\maketitle
\begin{abstract}
In this paper, we presented a detailed timing analysis of a prominent outburst of 4U 0115+63 detected by \textit{Insight}-HXMT in 2017 August. The spin period of the neutron star was determined to be $3.61398\pm 0.00002$ s at MJD 57978. We measured the period variability and extract the orbital elements of the binary system. The angle of periastron evolved with a rate of $0.048\pm0.003$ $yr^{-1}$. The light curves are folded to sketch the pulse profiles in different energy ranges. A multi-peak structure in 1-10 keV is clearly illustrated. We introduced wavelet analysis into our data analysis procedures to study QPO signals and perform a detailed wavelet analysis in many different energy ranges. Through the wavelet spectra, we report the discovery of a QPO at the frequency $\sim 10$ mHz. In addition, the X-ray light curves showed multiple QPOs in the period of $\sim 16-32 $ s and $\sim 67- 200 $ s. We found that the $\sim100$ s QPO was significant in most of the observations and energies. There exist positive relations between X-ray luminosity and their Q-factors and S-factors, while the QPO periods have no correlation with X-ray luminosity. In wavelet phase maps, we found that the pulse phase of $\sim 67- 200 $ s QPO drifting frequently while the $\sim 16-32 $ s QPO scarcely drifting. The dissipation of oscillations from high energy to low energy was also observed. These features of QPOs in 4U 0115+63 provide new challenge to our understanding of their physical origins.

\end{abstract}

\begin{keywords}
stars: individual: 4U 0115+63 -- X-rays: binaries -- X-rays: stars
\end{keywords}



\section{Introduction}

First discovered by the Uhuru satellite \citep{giacconi1972}, 4U 0115+63 is one of the best studied Be/X-ray binary in high energy astrophysics, which consists of a neutron star and a Be star. Precise positional determinations have been conducted by \textit{SAS-3}, \textit{Ariel V} and \textit{HEAO-1}, by which the companion star was successfully identified as a reddened O9e star, V635 Cas, with visual magnitude $V\approx 15.5$ \citep{cominsky1978,rap1978,hutching1981}. The distance of the star is determined to be about 7 kpc \citep{Negueruela2001}. Given the B0.2Ve spectral type of V635 Cas, its mass is expected to be about $19 M_{\sun}$ \citep{vacca1996,Negueruela2001}. The  neutron star has a $\sim3.6$ s spin period and its orbital elements were measured continuously \citep{cominsky1978, rap1978}. By \textit{SAS-3}, early research found the eccentric to be about 0.34 and orbital period being 24.3d respectively. Timing analysis showed an upper limit on the apsidal motion, being $\dot{\omega}\le 2.1^{\circ}$ yr$^{-1}$ \citep{kelley1981}.

In most of the time, a neutron star in a Be/X-ray binary will stay away from the Be circumstellar disc, but during the periastron passage, an abrupt mass accretion may take place, resulting a Type I outburst. Having duration of a few days to a few tens of days, peak X-ray luminosity during these outbursts is between $10^{35}$ to $10^{37} \rm\, erg\,s^{-1}$. Occasionally, Be/X-ray binaries (BeXBs) show giant outbursts, known as Type II outbursts, which is quite irregular and not linked with the binary orbit. These are caused by the enhanced episodic outflow of the Be star. Luminosity of these outbursts peak at $L_x\geqslant 10^{37} \rm\, erg\,s^{-1}$ \citep{boldin2013}. The neutron star in 4U 0115+63 shows Type II outbursts in a period of three to four years. \citet{Negueruela2001} explained this quasi-periodic behaviour as arising from the loss and reformation of a circumstellar disc around the Be companion star V635 Cas. They also suggested that the disc of Be/X-ray binaries is tidally truncated, resulting in the long quiescence period and lack of Type I outburst of this system.

The continuum spectrum of this pulsar was well described by a power-law model with high-energy cut-off \citep{white1983}. Research on cyclotron resonance spectral features (CRSFs) of this source has discovered five cyclotron line harmonics at $\sim11.2$, 22.9, 32.6, 40.8 and 54 keV using \textit{RXTE} \citep{ferrigno2009}. \citet{white1983} suggested that the $\sim20$ keV CRSF was the second harmonic resonance. These lines were confirmed by \citet{boldin2013} at $\sim$11, 24, 35.6, 48.8 and 60.7 keV by different continuum models. Using 11 keV as the fundamental line energy, the magnetic field of the neutron star in the absorption region was inferred to be $\sim10^{12}$ G. The luminosity dependence of the fundamental absorption-line in 4U 0115+63 was reported by \citet{nakajima2006} in 1999 outburst with \textit{RXTE}. They interpreted the anti-correlation between the CRSF energy and X-ray luminosity as a consequence of the decrease in height of the accretion column.

Except for the 3.61s pulsations induced by the neutron star spin, multiple quasi-periodic oscillations (QPOs) have been detected in its light curves. QPOs varying in the $\sim 27-46 $ mHz range were detected by \textit{Rossi X-ray Timing Explorer(RXTE)/Proportional Counter Array (PCA)} observations during the 1999, 2004 and 2008 outbursts \citep[see][]{dugair2013}. Later, with \textit{RXTE}, a $\sim2$ mHz QPO was reported by  \citet{heindl1999}. \citet{roy2019} with \textit{LAXPC/AstroSat} in 2015 outburst confirmed it and detected another $~\sim 1$ mHz QPO from \textit{Large Area X-ray Proportional Counter (LAXPC)} observations of 4U 0115+63 on 2015 October 24 .

With its high sensitivity and timing resolution from 1-250 keV, \textit{Insight}-HXMT was managed to observe a Type II outburst in 2017 ($8\times10^{36}\, {\rm \;erg/s}\lessapprox L_x\lessapprox 7\times 10^{37}\;{\rm erg/s}$). In this study, observations from \textit{Insight}-HXMT would be analyzed extensively. Archival data of 4U 0115+63 during the outburst in 2017 are summarized in Table~\ref{tab:obs}. Though temporal properties would be the concentration of this paper, a schematic spectral analysis is also carried to estimate X-ray flux of the source.

This paper is organized as followed. In \S 2, the observations of \textit{Insight}-HXMT on 4U 0115+63 are introduced, the analysis processes of the raw data are also presented. In \S 3, the science data analysis and results are presented, including the pulse period search, pulse profiles and orbital elements, spectral analysis and source fluxes, the QPO properties. We discussed the physical origins of the observed QPOs in \S 4. The conclusion is shown in \S 5.

\section{OBSERVATIONS}
Launched on June 15, 2017, \textit{Insight}-HXMT is the first X-ray astronomy satellite of China \citep{zhang2020}. HXMT operates in a low earth orbit and a total of three main payload is carried, which are the High Energy X-ray telescope (HE, from 20-250 keV, \citet{liu2020}), the Medium Energy Telescope (Me, 5-30 keV, \citet{cao2020}) and the Low Energy telescope (LE, 1-15 keV, \citet{chen2020}), having detector areas of 5100, 952 and 384 $\rm cm^2$  respectively. The time resolutions are 1 ms for LE, 280 $\rm\mu s$ for ME and 25 $\rm\mu s$ for HE. HE telescope is made up of 18 cylindrical NAI (TI)/CSI (Na) detector, each having a diameter of 190 mm and thickness of 3.5 mm and 40 mm for NaI and CsI respectively.

\textit{Insight}-HXMT Data Analysis software (HXMTDAS) V2.02 is used to analyze the observations. Using tasks \textit{hegtigen, megtigen} and \textit{legtigen}, data are filtered with criteria: (1)  pointing offset angle < 0.04$^\circ$; (2)  elevation angle > 10$^\circ$; (3)  geomagnetic cutoff rigidity > 8 GeV. Background light curves are estimated by making use of the linear correlation between the small filed of view detector and blind detectors, while the coefficient is calculated from the ratio of the number of non-blind detectors to that of blind detectors. \textit{Insight}-HXMT background team have tested this method in blank sky observations. Other researchers have adopted it successfully in the timing analysis of MAXI J1535-571 \citep{J1535} and MAXI J1820+070 \citep{J1820}, where the background systematic errors are 3.2\% (LE), ~3\% (ME) and 2.0\% (HE) \citep[see][]{HELEback,MEback}. Light curves are then generated using the standard HXMTDAS task \textit{"he/me/lelcgen"} (0.078125 s binned).

\begin{figure*}
	\centering
	\includegraphics[height=8cm]{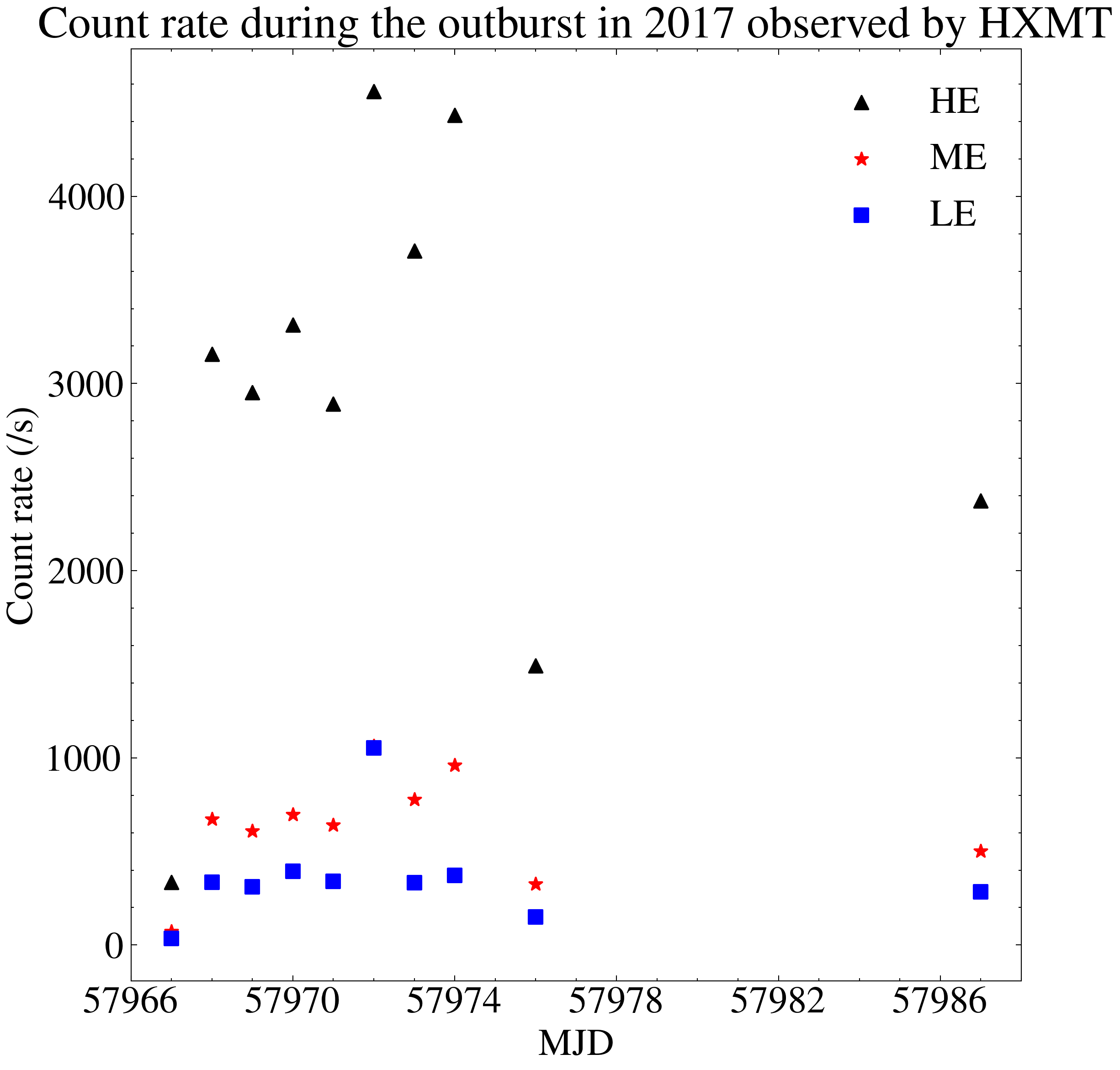}
    \includegraphics[height=8cm]{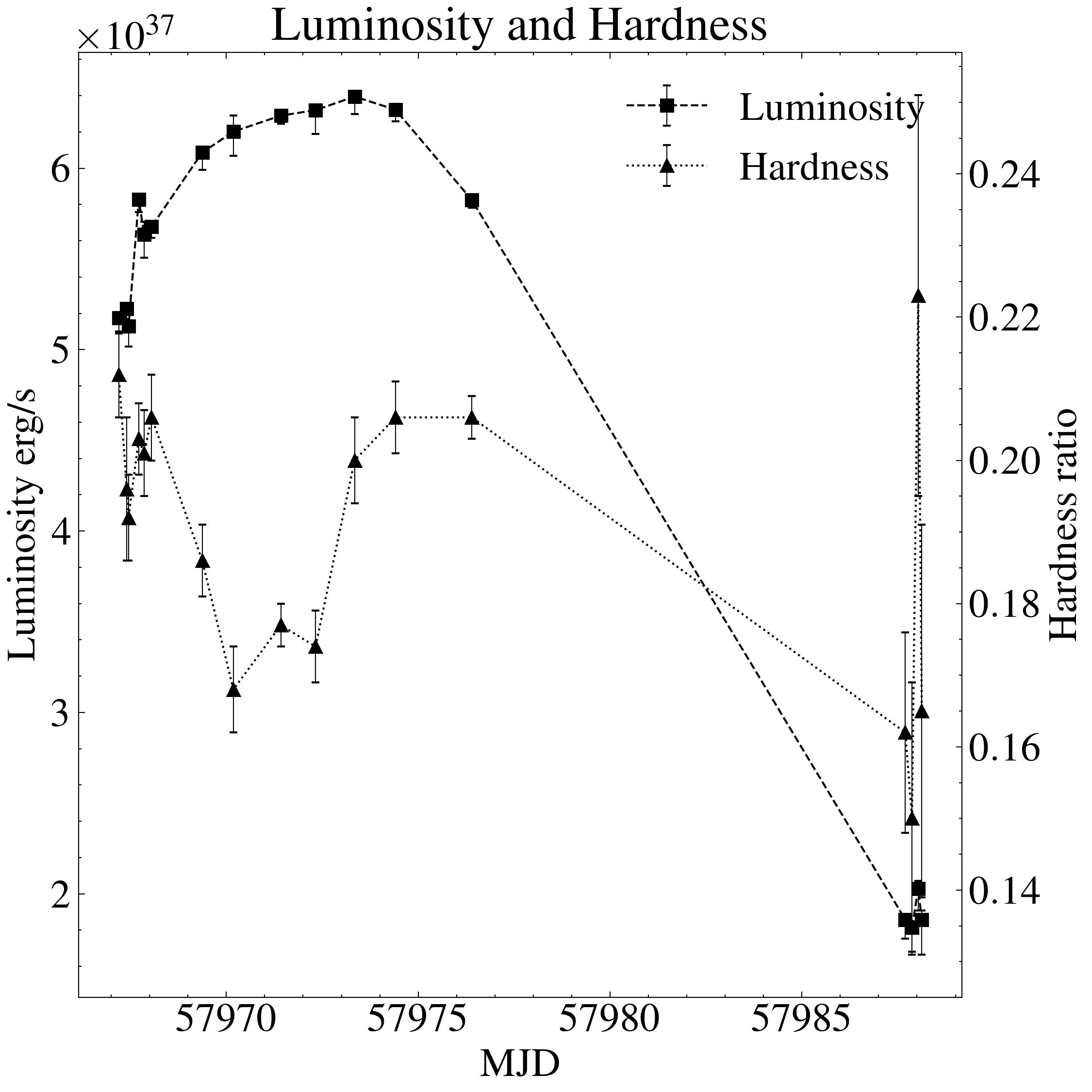}
	\caption{{\bf Left} Average photon counts per second in 2017 outburst. Data were obtained from \textit{Insight}-HXMT. Counts from different telescopes (HE, ME, LE)  have been averaged at day. It is clear to see that the outburst came to its climax at about MJD 57974 and then the source went dimmed very quickly in only a few days. {\bf Right:} Luminosity and hardness (30-50 keV over 3-10 keV) history of 4U 0115+63 observed by \textit{Insight}-HXMT. Peak luminosity during 2017 outburst occurred in MJD 57974, slightly shifted from the peak of X-ray hardness.}
	\label{fig:countrate}
\end{figure*}

The count rates and hardness ratio of 4U 0115+63 versus time have been shown in Figure~\ref{fig:countrate}, where the hardness is represented by the ratio of flux in 30-50 keV to that in 3-10 keV.  For the timing analysis, we adopt no dead correction because it will only affect white noise level and the normal dead time of \textit{Insight}-HXMT telescopes are in $\rm\mu s$ magnitude, which is far smaller than the oscillations and periods we are searching for. Moreover, a white noise background spectrum will be compared to our result so as to generate significance levels, by which we will be able to tell the true signal.

\begin{table}
	\centering
	\caption{Log of Observations of 4U 0115+63 for \textit{Insight}-HXMT.}
	\label{tab:obs}
		
		\begin{tabular}{lccc} 
			\hline
			Observations &Duration (s)& Time of Observation (UTC)&MJD\\
			\hline
			P0111517001&81211 &2017-08-02 04:03:10&57967\\
			P0111517002& 12625 &2017-08-04 08:34:28&57969\\
			P0111517003&12039&2017-08-05 03:40:20&57970\\
			P0111517004& 12039&2017-08-06 09:54:30&57971\\
			P0111517005&12460&2017-08-07 06:35:43&57972\\
			P0111517006&12528&2017-08-08 08:03:19&57973\\
			P0111517007&12268&2017-08-09 09:30:51&57974\\
			P0111517008&46915&2017-08-11 09:14:53&57976\\
			P0111517011&46879&2017-08-22 15:47:23&57988\\
			\hline
		\end{tabular}
\end{table}

\section{DATA ANALYSIS AND RESULTS}
\subsection{Pulse profiles and orbital elements}
\label{orbital elements}
Almost all X-ray sources in the universe are variable, with their intensity or flux changeing with time. These changes can be either periodic or quasi-periodic. Timing analysis is the study of such variability of X-ray sources. One of the main goals for timing analysis is to find out underlying physical mechanism of any variability.

Pulse profile variations in X-ray binaries have not been well understood. Thanks to the wide energy range of \textit{Insight}-HXMT, it is achievable to divide time series into different energy bands, which are 1-10, 10-20, 20-30, 30-50, 50-70 and 70-100 keV respectively. In this study, we are going to investigate the evolution of pulse profiles with different energy ranges.

Orbital elements of an X-ray binary system are an important tool for unveiling the outburst mechanism. In such an accretion system, the apparent pulsed period changes continuously because it is highly sensitive to binary orbital elements. So, we will try to derive orbital parameters from observed apparent spin periods. Besides, as we have derived the angle of periastron, we studied the apsidal motion of 4U 0115+63 by comparing our result with previous researchers' work.

To perform these analysis, we should first obtain the apparent pulse period using the Insight-HXMT data. Traditionally, one may use pulse arrival time to determine pulse period of X-ray pulsars in binaries \citep{rap1978}. Pulse arrival time for each data was determined by comparing folded pulse profiles with specific reference profile. However, the pulse profile of the neutron star in a binary system is very likely to change in the course of a giant outburst, so it seems inappropriate to measure its period by this method. In our work, we adopt the method first used by \citet{raichur2010} , which was also followed by \citet{jli2012}. This method is simpler than the previous one, and would not be affected by the variation of pulse profiles.

Since the determination of the spin period is significantly affected by count rates of the observed source, it is reasonable to stack all HE (pulsations in high energy bands are more likely to be a single peaked sinusoidal shape) counts together, by which we can reduce the error effectively. The arrival time of each photon is corrected to the barycentre of the Solar-system by HXMTDAS task \textit{hxbary}. After that, we used HEASoft, a software package published by NASA, and used python library Astropy to handle the generated FITS files, and reprocess them to the highest level scientific products. In this procedure, we performed a folding analysis in the time domain using HEASoft built-in function \textit{efsearch} to calculate the $\chi^{2}$ and fitted the distribution with the theoretical distribution function given by \citet{leahy1987}. We have noticed that the sampling rate of our data was uneven. Consequently, the error was estimated from Equation 2 of \citet{larsson1996}, where the parameter \textit{a} was taken to be 0.469.

Next, we started to determine the orbital elements. As given by \citet{raichur2010}, in a relatively short time elapse (for which we can consider the first derivative of spin period as constant), there is an approximated relation of Doppler modulated and observed period of an X-ray source:
\begin{equation}
    P^{\rm obs}_{\rm spin}(t)=[P_{\rm spin}(t_0)+(t-t_0)\dot{P}_{\rm spin}]\sqrt{\frac{1+v_r/c}{1-v_r/c}}.
	\label{eq:spinperiod}
\end{equation}
Here, $\dot{P}_{\rm spin}$ is the derivative of spin period, while observed spin period at time $t$ is denoted by $P^{\rm obs}_{\rm spin}(t)$. $t_0$ is a reference time of the outburst. $v_r$ is radial velocity of the source. It is the component of velocity along the line of sight and observes the following equation:

\begin{equation}
v_r=\frac{2\pi a_x \sin i}{(1-e^2)^{1/2}P_{\rm orb}}[\cos(\nu+\omega)+e\cos\omega]
\end{equation}
Projected semi-major axis is denoted by $ \rm a_x\sin i$ here and $\rm \omega$ is the angle of periastron. \textit{e} is orbital eccentricity while $\nu$ is the true anomaly. Their relations are given as follows:

\begin{equation}
	\tan \frac{\nu}{2}=\tan \frac{E}{2}\sqrt{\frac{1+e}{1-e}}.
\end{equation}

\begin{equation}
E-e\sin E=\frac{2\pi}{P_{\rm orb}}(t-T_{\rm\omega}).
\label{eqn:E}
\end{equation}

E in equation~\ref{eqn:E} denotes the eccentric anomaly. Combine these equations together and make use of observed spin periods, we can solve for the Doppler motion, thus obtaining orbital elements.

\begin{figure}
	\includegraphics[width=\columnwidth]{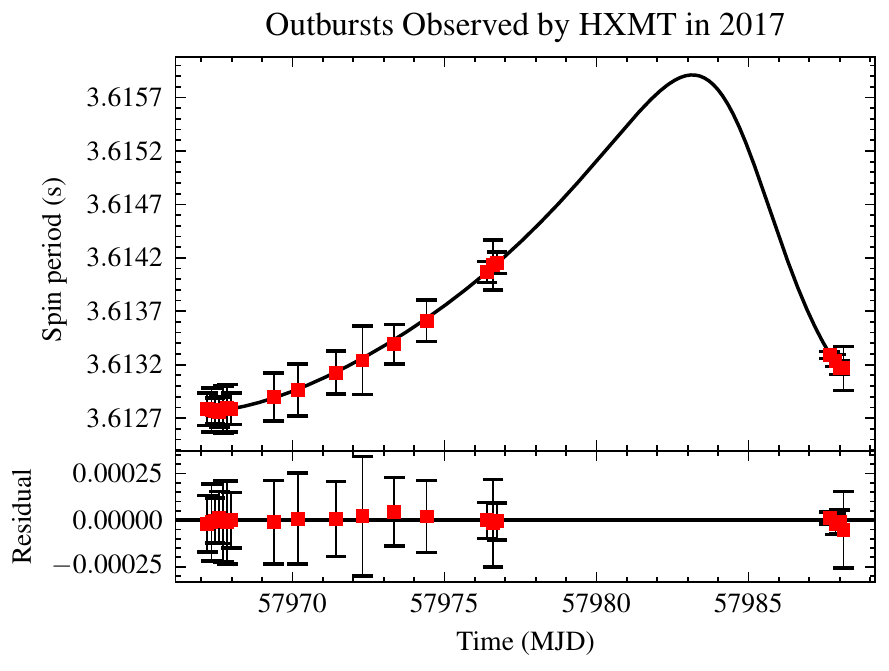}
	\caption{Fitted spin period of neutron star in 4U 0115+63 during the 2017 outburst. All data are from \textit{Insight}-HXMT. Residuals are shown in the lower panel. Periods of this neutron star are modulated by Doppler motion. Solid curve represents the predicted Doppler motion of the binary as a consequence of the best-fitting orbital elements.}
	
	\label{fig:period}
\end{figure}

We fit the observed spin period values in different time intervals during the 2017 outburst. All the observation data points are presented in Figure~\ref{fig:period}. This fitting is based on the Levenberg-Marquardt algorithm, an algorithm designed for non-linear optimization, realized by Scipy built-in function \textit{leastsq}. Every parameter's error was calculated from the covariance matrix of its estimate. Some of the parameters were adopted from \citet{bildsten1997}, like the orbital period $P_{\rm orb}$ and projected semi-major axis. Results are shown in Table~\ref{tab:elements}.

As shown in Table~\ref{tab:pdot} and \ref{tab:omega}, the parameters are generally consistent with previous results in the outburst history of the source. While the neutron star in 4U 0115+63 is still spinning up during this outburst, it had a relatively smaller derivative than that was observed in 2008. However, due to the lack of observation between MJD 57978 and MJD 57986, the uncertainty of spin period derivative cannot be well constrained.

\begin{figure}
	\includegraphics[width=\columnwidth]{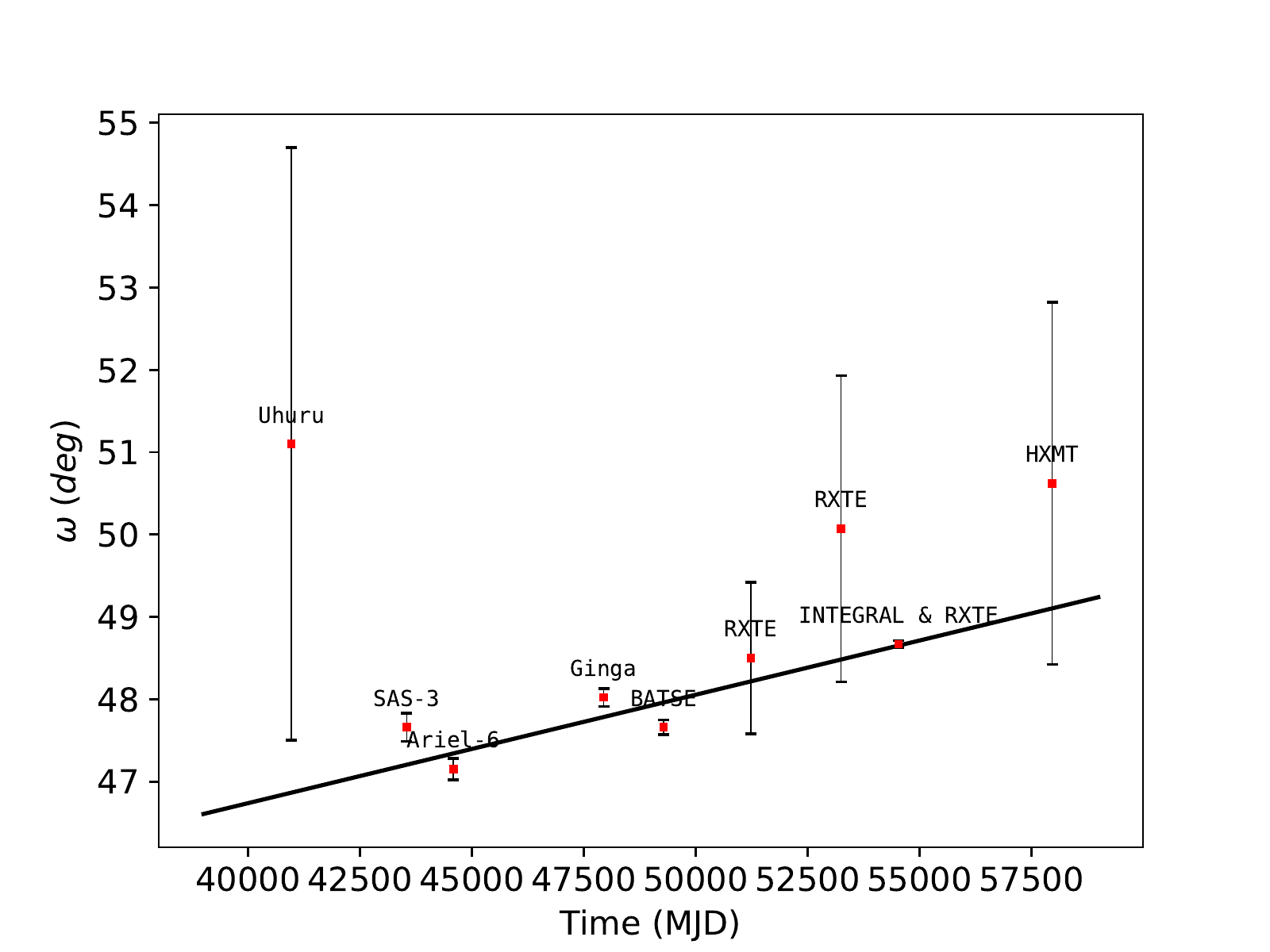}
	\caption{Measured $\omega$ values of 4U 0115+63 from 1971 to 2017. A linear fitting gives the average $\dot{\omega}=0.048^\circ \pm0.003^\circ \rm\, /yr$}.
	\label{fig:apsidal}
\end{figure}

Considering the $ \rm \omega$ values estimated by previous researchers in Table \ref{tab:omega}, we are able to measure the apsidal motion of this binary system. Since we have derived the angle of periastron $ \rm\omega=50.62^\circ \pm2.20^\circ $ in 2017, it is possible to combine it with the previous measurements and then generate the corresponding apsidal motion rate. Relevant data and fitted line are plotted in Figure~\ref{fig:apsidal}. Using the same method, the motion rate is determined to be $ \dot{\omega}=0.048^\circ \pm0.003^\circ \rm\, /yr $; error is calculated from the covariance matrix either.

\begin{figure*}
	\centering
	
	\subfigure{
		\begin{minipage}[t]{0.4\linewidth}
			\centering
			\includegraphics[width=7cm]{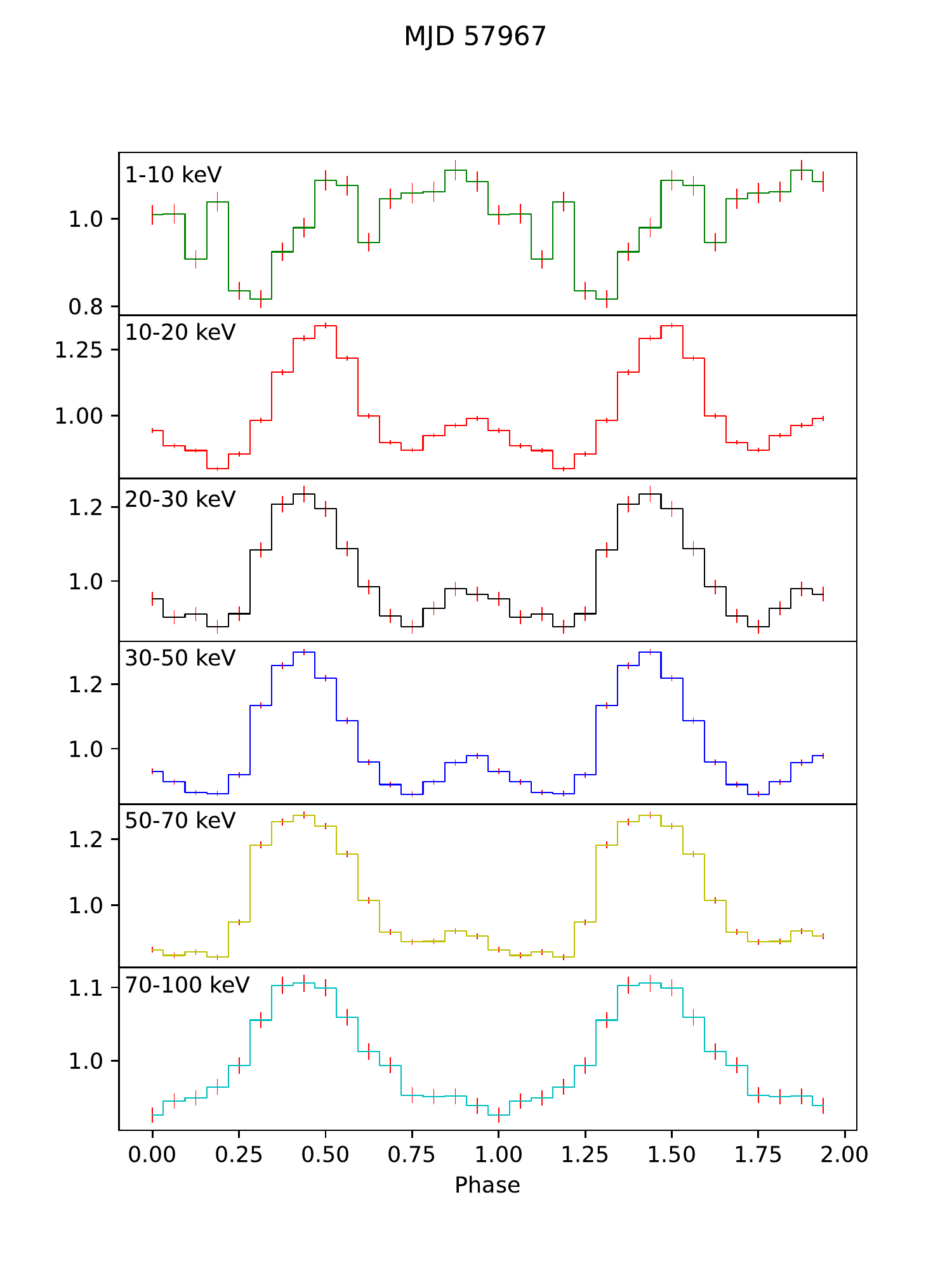}
		\end{minipage}%
	}%
	\subfigure{
		\begin{minipage}[t]{0.4\linewidth}
			\centering
			\includegraphics[width=7cm]{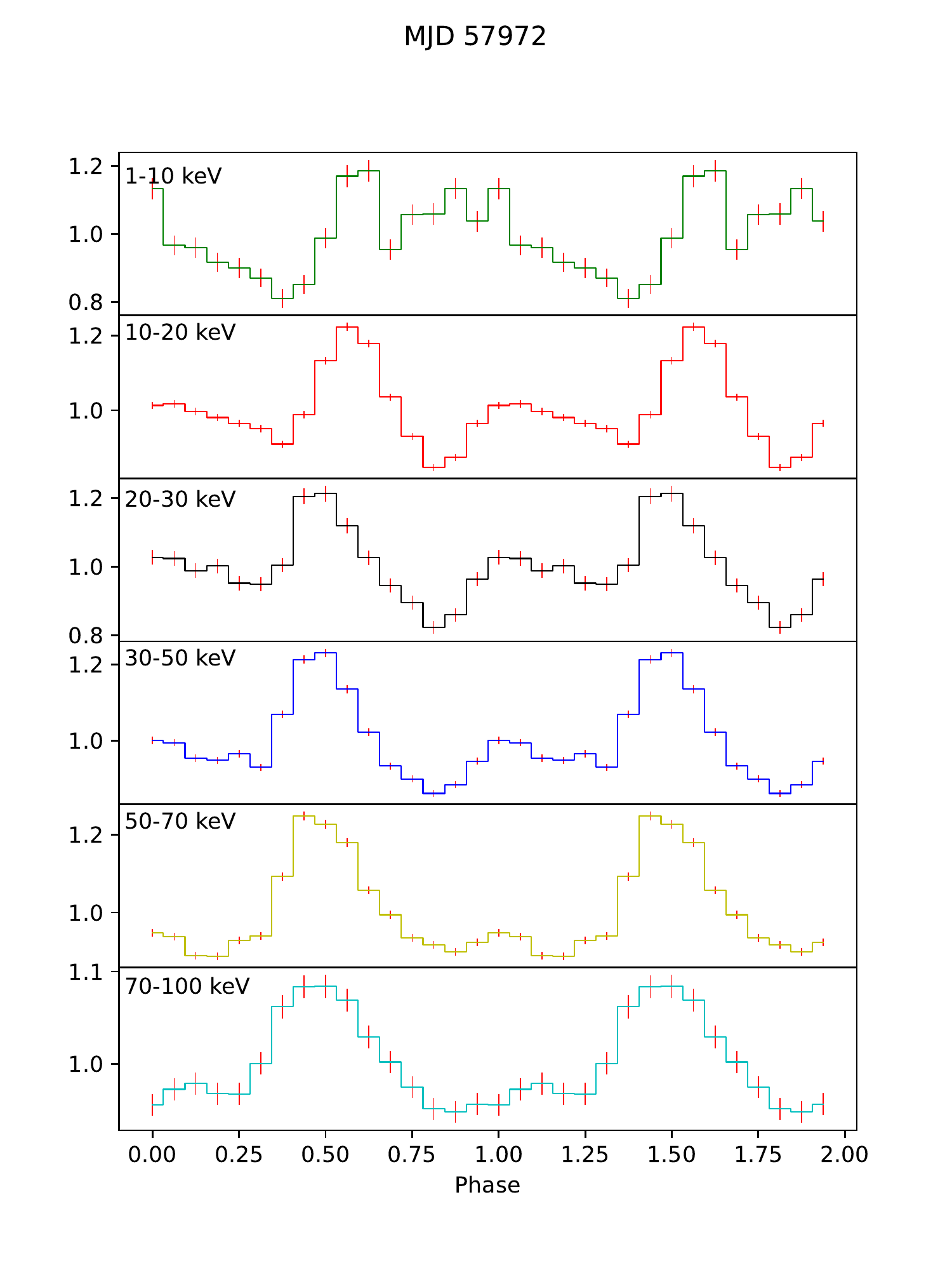}
		\end{minipage}%
	}%
	
	\subfigure{
		\begin{minipage}[t]{0.4\linewidth}
			\centering
			\includegraphics[width=7cm]{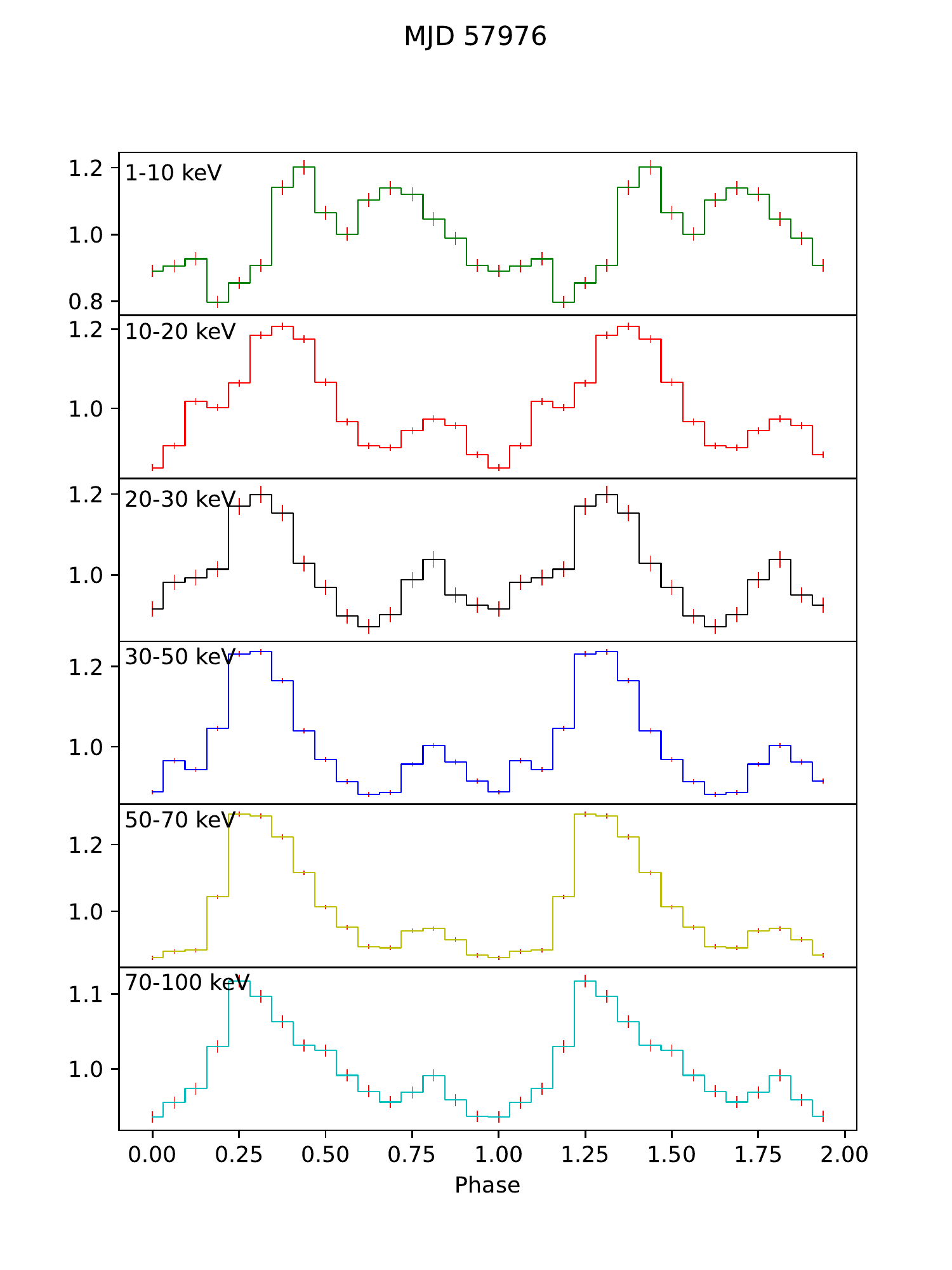}
		\end{minipage}
	}%
	\subfigure{
		\begin{minipage}[t]{0.4\linewidth}
			\centering
			\includegraphics[width=7cm]{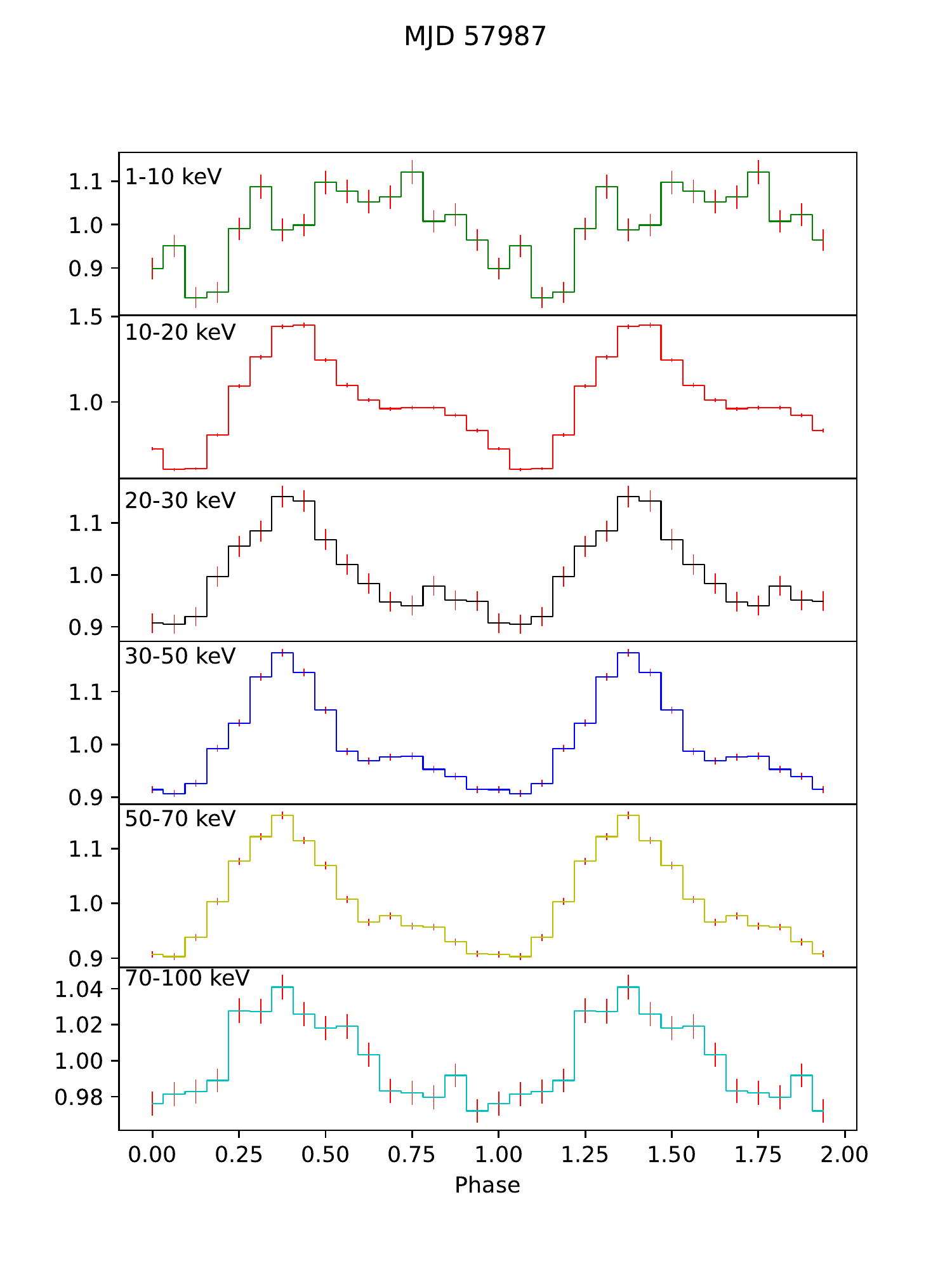}
		\end{minipage}
	}%
	
	\centering
	\caption{Pulse profiles of 4U 0115+63 during 2017 outburst. Second peak dissipated as the outburst draw to an end. In the band of 1-10 keV, the source shows the multiple peaks, while in the higher energies, shows double peak features, and the mini peak disappears above 70 keV.}	
	\label{fig:pprofile}
\end{figure*}

In addition, we studied the evolution of pulse profiles in 4U 0115+63. We derived the pulse profiles from pulse periods by epoch folding. Here, we used \textit{efold} function and reprocessed generated FITS files by python to produce fancy graphics. We chose the time resolution as 0.078125 s, which is sufficiently small to analyze this source while being large enough to avoid most of the statistical errors and noises. The error bars are evaluated by propagating the theoretical error bars through an averaging process. We illustrate pulse profiles both with the outburst time and energy ranges. It is worth noting that the count rate above 100 keV becomes too small, relative errors for light curves folding surge rapidly, so we only analyze the data lower than 100 keV. Results are plotted in Figure~\ref{fig:pprofile}. There are very prominent second peaks in most of the time of the outburst. We reconfirmed the phenomenon discovered by \citet{jli2012} that the secondary peak's amplitude have positive correlation with the outburst flux. As the flux arrives at its peak, the secondary peak also starts to merge with the main peak. Moreover, we found that generally, the higher the energy, the lower its peak amplitude. Pulse profiles of 1-10 keV are clearly very different from others, which shows the multiple peak features.

\begin{table}
	\centering
	\caption{Spin period and orbital elements of 4U 0115+63 estimated from 2017 outburst.}
	\label{tab:elements}
	\begin{threeparttable}

		\begin{tabular}{lc} 
			\hline
			Parameter & Value \\
			\hline
            $P_{\rm spin}$ (s, at MJD 57977) &   3.61398$\pm 0.00002$\\
			$\dot{P}_{\rm spin}$ (s\ d$^{-1})$ & $(-4.26 \pm 6.21)\times 10^{-6}$ \\
			$\omega\;(\deg)$ & $50.62 \pm 2.20$ \\
			$\dot{\rm\omega}\;(\deg\rm/yr)$&$0.048\pm0.003$\\
			$T_\omega$ (MJD) & $57960.543 \pm 0.063$ \\
			$P_{\textit{\rm orb}}$ (d)\tnote{$a$} &$24.317037\pm 0.000062$ \\
			$e$\tnote{$a$} & $0.3402\pm 0.0002$\\
			$\textit{a}_{x}\sin i$ (light-second)\tnote{$a$}&$140.13\pm 0.08$\\
			\hline
		\end{tabular}
		\begin{tablenotes}
			\footnotesize
			\item[$a$] By referring the results of \citet{bildsten1997}
		\end{tablenotes}
	\end{threeparttable}
\end{table}

\begin{table*}
	\centering
	\caption{History of intrinsic spin period and its derivative of 4U 0115+63.}
	\label{tab:pdot}
	\begin{threeparttable}
		\begin{tabular}{cccc}
			\hline
			MJD & $P_{\rm spin}$ (s) & $\dot{P}/P_{\rm spin}$ (yr$^{-1}$) & References\\
			\hline
			40963&$3.614658\pm 0.000036$&$(-3.3\pm1.4)\times 10^{-6}$ &\citet{kelley1981}\\
			42283&$3.6142\pm 0.0001$&$-3.4\times10^{-5}$\tnote{$a$}&\citet{whitlock1989}\\
			43540&$3.6145737\pm 0.0000009$&$(-3.2\pm0.8)\times 10^{-5}$&\citet{rap1978}\\
			44589&$3.6146643\pm 0.0000018$&$(-2.6\pm0.5)\times10^{-4}$&\citet{ricketts1981}\\
			47941&$3.614690\pm 0.000002$&$(-1.8\pm0.2)\times 10^{-4}$&\citet{tamura1992}\\
			49481&$3.6145107\pm 0.0000010$&$(-1.8\pm0.6)\times 10^{-4}$&\citet{scott1994}\\
			50042&$3.614499\pm 0.000004$&$(-4.2\pm0.9)\times 10^{-5}$&\citet{finger1995}\\
			50307&$3.614451\pm 0.000006$&-&\citet{scott1994}\\
			51232&$3.614523\pm 0.000003$&-&\citet{wilson1999}\\
			51240&$3.61447 \pm 0.00002$&$(-2.52\pm0.60)\times 10^{-4}$&\citet{raichur2010}\\
			53254&$3.61436 \pm 0.00002$&$(-3.84\pm0.55)\times 10^{-4}$&\citet{raichur2010}\\
			54566&$3.61430 \pm 0.00002$&$(-7.31\pm0.03)\times 10^{-4}$&\citet{jli2012}\\
			57977&$3.61398 \pm 0.00002$&$(-4.3\pm6.3)\times 10^{-4}$&this work\\
			\hline
		\end{tabular}
		\begin{tablenotes}
			\footnotesize
			\item[$a$] By directly comparing previous spin period in 1971 and 1974.
		\end{tablenotes}
	\end{threeparttable}
\end{table*}

\begin{table*}
	\centering
	\caption{Epoch and angle of periastron history of 4U 0115+63.}
	\label{tab:omega}
	\begin{threeparttable}
		\begin{tabular}{ccccc}
			\hline
			Orbit number & Periastron time passage(TJD) & $\omega(\deg)$ & Satellite & References\\
			\hline
			0&$40963.08 \pm 0.17$&$51.10\pm3.60$ & \textit{Uhuru} &\citet{kelley1981}\\
			106&$43540.451 \pm 0.006$&$47.66\pm0.17$&\textit{SAS-3}&\citet{rap1978}\\
			149&$44585.700$\tnote{$a$}&$47.15\pm0.13$&\textit{Ariel-6}&\citet{ricketts1981}\\
			287&$47941.530 \pm 0.006$&$48.02\pm0.11$&\textit{Ginga}&\citet{tamura1992}\\
			342&$49279.2677 \pm 0.0034$&$47.66\pm0.09$&\textit{BATSE}&\citet{cominsky1978}\\
			422&$51224.6465 \pm 0.051$&$48.50\pm0.92$&\textit{RXTE}&\citet{raichur2010}\\
			505&$53243.038\pm 0.051$&$50.07\pm1.86$&\textit{RXTE}&\citet{raichur2010}\\
			558&$54531.7709 \pm 0.0603$&$48.67\pm0.04$&\textit{INTEGRAL} and \textit{RXTE}& \citet{jli2012}\\
			697&$57960.543 \pm 0.063$&$50.62\pm2.20$&\textit{Insight}-HXMT&this work\\
			\hline
			
		\end{tabular}
		\begin{tablenotes}
			\footnotesize
			\item[$a$] Derived from \citet{ricketts1981}.
		\end{tablenotes}
	\end{threeparttable}
\end{table*}

\subsection{Spectral analysis}

\begin{figure}
	\includegraphics[width=6cm,angle=-90]{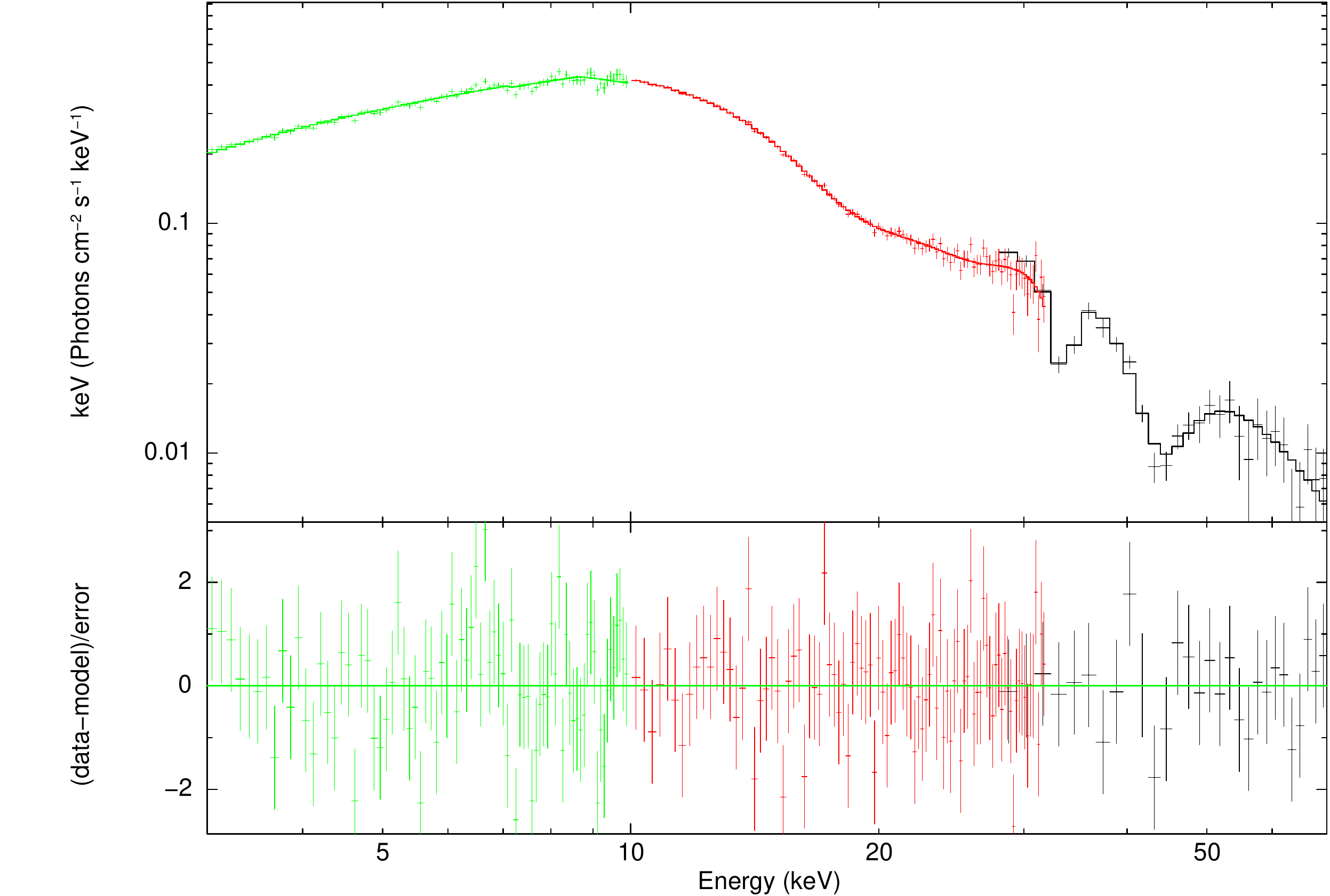}
	\vspace{0.8cm}
	\caption{Fitted spectra of 4U 0115+63 in MJD 57968 with the energy range from 3 to 70 keV. HE (28-70 keV), ME (10-32 keV) and LE (3-10 keV) are denoted by black, green and red data points respectively. See Table~\ref{tab:spec} for detailed parameter values.}
	
	\label{fig:spectrum}
\end{figure}

A schematic spectral analysis have been performed to obtain the absolute X-ray luminosity of the source during the 2017 outburst. Data from three telescopes of \textit{Insight}-HXMT were merged to derive the source spectrum. We fitted four cyclotron absorption lines for this sources based on the \textit{Insight}-HXMT data. However, due to the relatively low signal to noise ratio in high energy, the width of some of these CRSFs was fixed (see Table~\ref{tab:spec}). A constant component was added to calibrate the unequal absolute flux between three telescopes. After that, we included interstellar absorption and a combined model of power law and high-energy cutoff \citep{boldin2013,white1983,tsygankov2007,wilms2000} :

\begin{equation}
f(E)=KE^{-\Gamma}\times
\begin{cases}
1&(E\le E_c)\\
\exp\{-(E-E_c)/E_f\}&(E>E_c),
\end{cases}
\end{equation}
where \textit{K} is the normalization factor, referring to photons $\rm keV^{-1}cm^{-2}s^{-1}$ at 1 keV while $\Gamma$ is the photon index of the power law. \textit{E} is photon energy and $E_c$ is the cutoff energy in keV, e-folding energy in keV is denoted by $E_f$. Then, cyclotron absorption line models in XSPEC are added to account for the spectrum structure at the four line energy bands: 11 $\sim$ 18 keV, $\sim24$ keV, $35\sim40$ keV, and $45\sim50$ keV \citep{nakajima2006,ferrigno2009,jli2012,boldin2013}. This model in XSPEC is defined as \citep{mihara1990}:

\begin{equation}
	M(E)=\exp\left[ -D_f\frac{(W_fE/E_{\rm cyc})^2}{(E-E_{\rm cyc})^2+W_f^2} \right],
\end{equation}
The depth of fundamental line is denoted by $D_f$ and its energy is $E_{\rm cyc}$ and $W_f$ is the width of it. Column density of interstellar medium is fixed at $N_H=1.3\times10^{22}\rm\;cm^{-2}$ \citep{iyer2015} for all observations. For a summary, the broad band X-ray spectra were fitted with a model of "constant*TBabs*powerlaw*highecut*cyclabs*cyclabs*cyclabs*cyclabs" (built in XSPEC). An example spectrum was presented in Figure~\ref{fig:spectrum}. Flux is estimated from the best-fit model using XSPEC built-in command "flux" in the 4-70 keV energy range. Assuming the distance to the neutron star \textit{d} to be 7 kpc, X-ray luminosity can then be calculated.

\begin{table*}
	\renewcommand\arraystretch{1.5}
	\centering
	\caption{Spectral parameters of 4U 0115+63 with \textit{Insigt}-HXMT in different time (MJD).}
	\label{tab:spec}
	\begin{threeparttable}
		
		\begin{tabular}{ccccccc} 
			\hline
			Model parameters &MJD 57967&MJD 57968&MJD 57971&MJD 57974&MJD 57977&MJD 57988\\
			\hline
			
			$E_{\rm cut}$ (keV)& $8.45^{+0.44}_{-0.32}$&$8.65^{+0.38}_{-0.34}$&$8.46^{+0.29}_{-0.27}$&$8.50_{-0.23}^{+0.25}$&$8.56_{-0.27}^{+0.33}$&$9.71^{+0.76}_{-0.71}$\\
			$E_{\rm fold}$ (keV)&$9.75^{+0.38}_{-0.38}$&$10.93^{+0.58}_{-0.57}$&$8.9_{-0.48}^{+0.45}$&$9.91^{+1.59}_{-0.40}$&$9.00_{-0.29}^{+1.09}$&$10.19^{+1.22}_{-1.26}$\\
			Photon index&$0.33^{+0.03}_{-0.02}$&$0.35^{+0.03}_{-0.03}$&$0.39^{+0.03}_{-0.03}$&$0.42_{-0.05}^{+0.02}$&$0.37_{-0.02}^{+0.02}$&$0.40^{+0.05}_{-0.04}$\\
			
			$\rm Depth_{cycl,1}$&$0.87^{+0.05}_{-0.09}$&$0.93^{+0.12}_{-0.13}$&$0.77_{-0.05}^{+0.06}$&$0.43^{+0.11}_{-0.26}$&$0.52^{+0.06}_{-0.14}$&$0.93_{-0.21}^{+0.14}$\\
			$E_{\rm cycl,1}$ (keV)&$18.21^{+0.20}_{-0.38}$&$17.64_{-0.35}^{+0.31}$&$17.18_{-0.11}^{+0.12}$&$16.69_{-0.45}^{+0.50}$&$17.62_{-0.62}^{+0.33}$&$17.47^{+0.40}_{-0.85}$\\
			$\rm Width_{cycl,1}$ (keV)&$6.29^{+0.45}_{-0.96}$&$5.93_{-0.90}^{+1.08}$&$4.55^{+0.54}_{-0.52}$&$4.02_{-1.08}^{+1.14}$&$4.42_{-0.97}^{+0.97}$&$7.11_{-1.76}^{+1.74}$\\
			
			$\rm Depth_{cycl,2}$&$0.55^{+0.20}_{-0.23}$&$0.37_{-0.18}^{+0.17}$&$0.31_{-0.10}^{+0.11}$&$0.64_{-0.11}^{+0.11}$&$0.48^{+0.18}_{-0.08}$&$0.92_{-0.59}^{+0.44}$\\
			$E_{\rm cycl,2}$ (keV)&$24.63^{+0.43}_{-0.43}$&$25.30^{+3.35}_{-1.60}$&$25.35_{-0.98}^{+0.96}$&$24.41_{-5.53}^{+1.19}$&$24.43^{+0.46}_{-0.96}$&$25.11_{-0.68}^{+1.03}$\\
			$\rm Width_{cycl,2}$ (keV)&$1.00^{+1.62}_{-1.00}$&5.00 (frozen)&2.00 (frozen)&$6.53_{-2.88}^{+8.22}$&$2.26_{-2.26}^{+4.01}$&$1.24^{+3.26}_{-1.24}$\\
			
			$\rm Depth_{cycl,3}$&$0.51^{+0.30}_{-0.24}$&$1.52_{-0.72}^{+0.87}$&$1.17_{-0.95}^{+0.86}$&$0.61^{+0.52}_{-0.61}$&$0.69^{+0.23}_{-0.37}$&$ 1.36^{+1.55}_{-1.06}$\\
			$E_{\rm cycl,3}$ (keV)&$31.14^{+0.97}_{-1.00}$&$33.35^{+1.21}_{-1.07}$&$32.29^{+1.22}_{-1.99}$&$32.48^{+7.45}_{-2.32}$&$30.07_{-0.77}^{+1.46}$&$33.08_{-2.12}^{+0.83}$\\
			$\rm Width_{cycl,3}$ (keV)&1.00 (frozen)&1.00 (frozen)&1.00 (frozen)&1.00 (frozen)&1.00 (frozen)&1.00 (frozen)\\
			
			$\rm Depth_{cycl,4}$&$1.12^{+0.20}_{-0.12}$&$1.52^{+0.31}_{-0.30}$&$0.64_{-0.30}^{+0.24}$&$0.95^{+0.21}_{-0.41}$&$0.71_{-0.16}^{+0.16}$&$1.38^{+0.83}_{-0.76}$\\
			$E_{\rm cycl,4}$ (keV)&$41.33^{+0.84}_{-0.86}$&$43.38^{+1.33}_{-1.34}$&$41.25^{+3.23}_{-4.99}$&$42.83^{+3.69}_{-1.29}$&$39.84_{-1.24}^{+1.55}$&$44.97^{+2.35}_{-2.63}$\\
			$\rm Width_{cycl,4}$ (keV)&5.00 (frozen)&5.00 (frozen)&5.00 (frozen)&5.00 (frozen)&5.00 (frozen)&5.00 (frozen)\\
			
			$\rm Flux_{4-70}\;(10^{-9}\;erg\;cm^{-2}s^{-1})$&8.75$^{+0.01}_{-0.86}$&9.69$^{+0.04}_{-0.12}$&10.73$^{+0.06}_{-0.08}$&10.78$^{+0.03}_{-0.17}$&9.94$^{+0.05}_{-0.08}$&3.17$^{+0.06}_{-0.21}$\\
			$\chi^2$ (d.o.f)&175.14 (187)&176.92 (188)&172.77 (187)&186.63 (187)&154.93 (187)&157.56 (187)\\
			Runs&0.25&-1.44&0.49&-0.19&0.38&0.36\\
			\hline
		\end{tabular}
	\end{threeparttable}
\end{table*}

\subsection{Wavelet analysis and QPOs}

In this section, we performed a wavelet analysis of light curves from a few keV up to 250 keV with \textit{Insight}-HXMT. Wavelet transform is a very convenient tool especially in non-stationary signal analysis. By decomposing time series into time-frequency space, one can easily determine both the dominant modes of variability and how those modes vary in time. This method has been widely applied in geophysics. Our procedure followed \citet{torrence1997}, here after TC98 (see Appendix~\ref{sec:wavelet} and TC98). The null hypothesis in our paper is generally the same as TC98. It is assumed that the time series has a mean power spectrum, which is a white noise one in this study. As a consequence, if a peak in the wavelet spectrum is above this background, we can assume the peak to be a true feature with a certain percent confidence. It should be noted that "significant at the 5\% level" is the same as "the 95\% confidence level", implying a test against a certain background level (see TC98 for more details). All QPO signals reported in this study are significant at the 5\% level.

To quantify the QPOs features globally, full width in half maximum (FWHM) and quality factor (Q-factor) are calculated and the deviations of global oscillation period (frequency) are estimated accordingly. To investigate  the stability of these oscillations, it may be wise to define a factor so as to make full use of the locality of wavelet transform. S-factor (stability factor) here is defined as the oscillation time (where (local) wavelet power in centroid period is greater than the 95\% significance spectrum) divided by the exposure time.

\begin{equation}
	S=\frac{oscillation\,time}{exposure\,time}
\end{equation}

\textit{R} factor is the power of the global wavelet spectrum relative to the global 95\% confidence spectrum, being used to filter the true features in this study.
\begin{equation}
	R=\frac{Global\,signal\,peak}{Global\,95\%\,confidence\,spectrum}
\end{equation}

We ruled out all peaks with \textit{R} factor less than 1. Global wavelet spectrum is the time-averaged one over all the local wavelet period, which is (TC98):
\begin{equation}
	\bar{W}(s)=\frac{1}{N}\sum_{n=0}^{N}|W_n(s)|^2,
\end{equation}

Q-factor has been calculated from FWHM (full width in half maximum):
\begin{equation}
	Q=\frac{Centroid\,frequency}{FWHM}
\end{equation}

\begin{figure*}
	\centering
	
	\subfigure[]{
		\begin{minipage}[t]{0.5\linewidth}
			\centering
			\includegraphics[width=9cm]{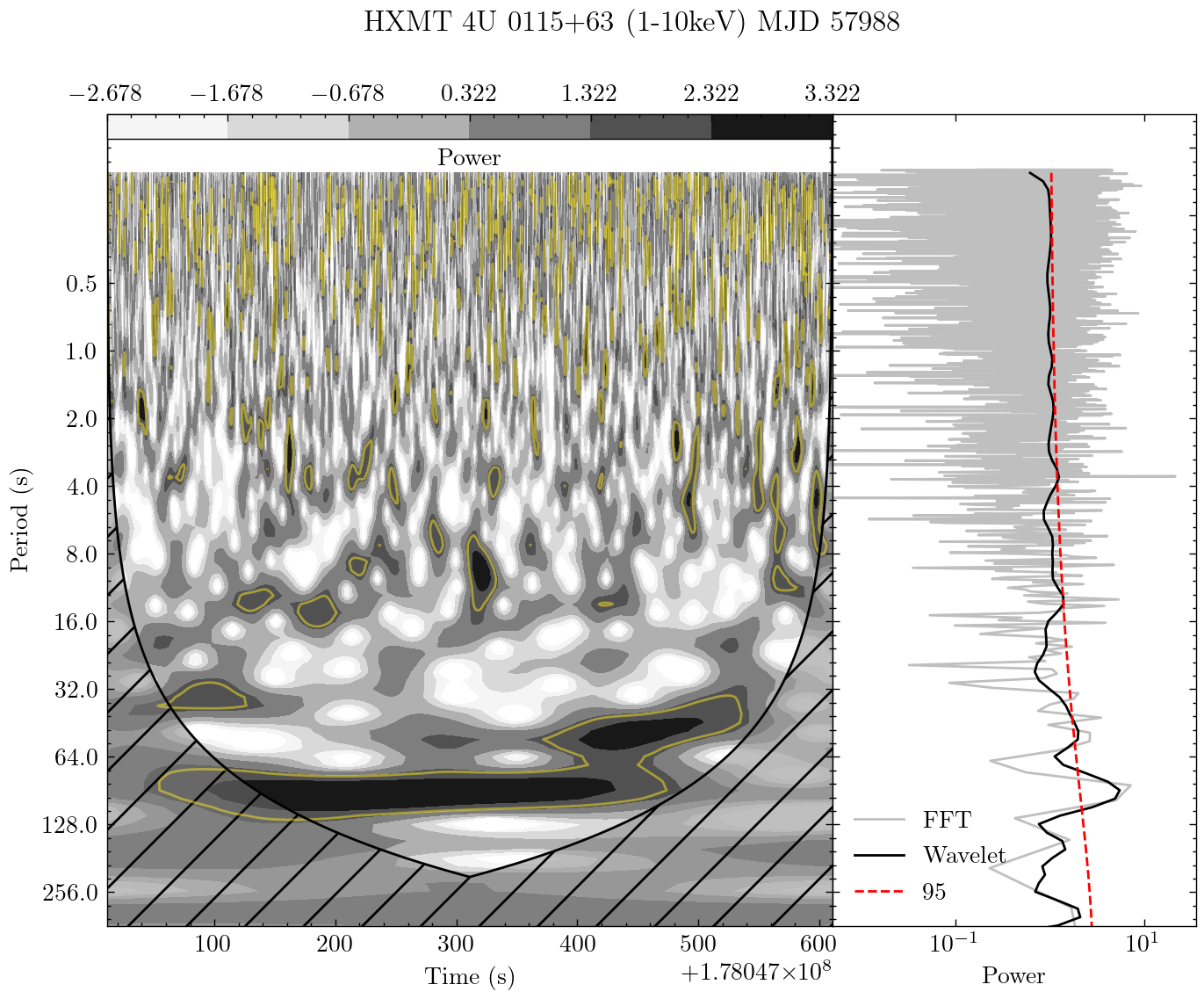}
		\end{minipage}%
	}%
	\subfigure[]{
		\begin{minipage}[t]{0.5\linewidth}
			\centering
			\includegraphics[width=9cm]{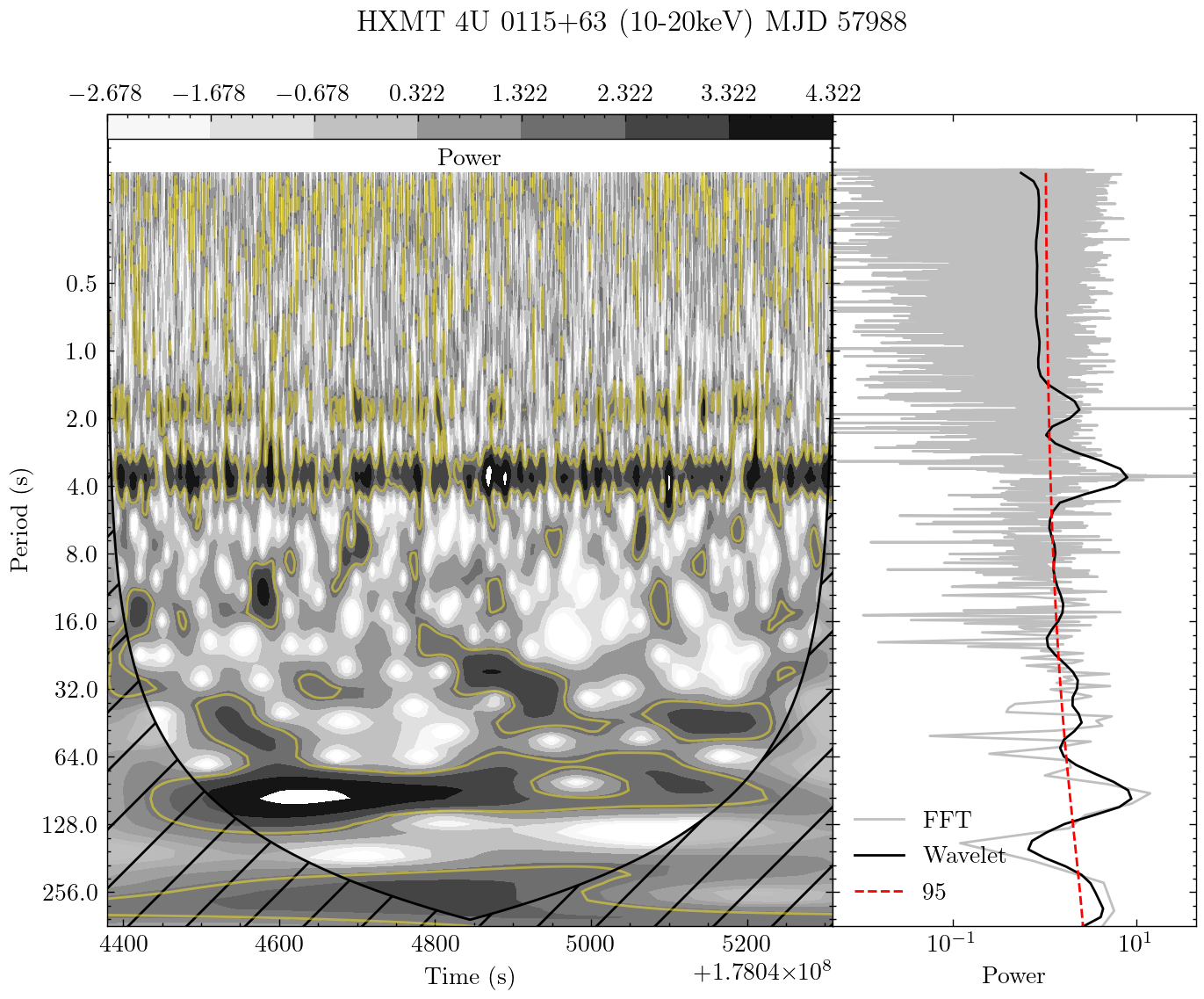}
		\end{minipage}%
	}%
	
	\subfigure[]{
		\begin{minipage}[t]{0.5\linewidth}
			\centering
			\includegraphics[width=9cm]{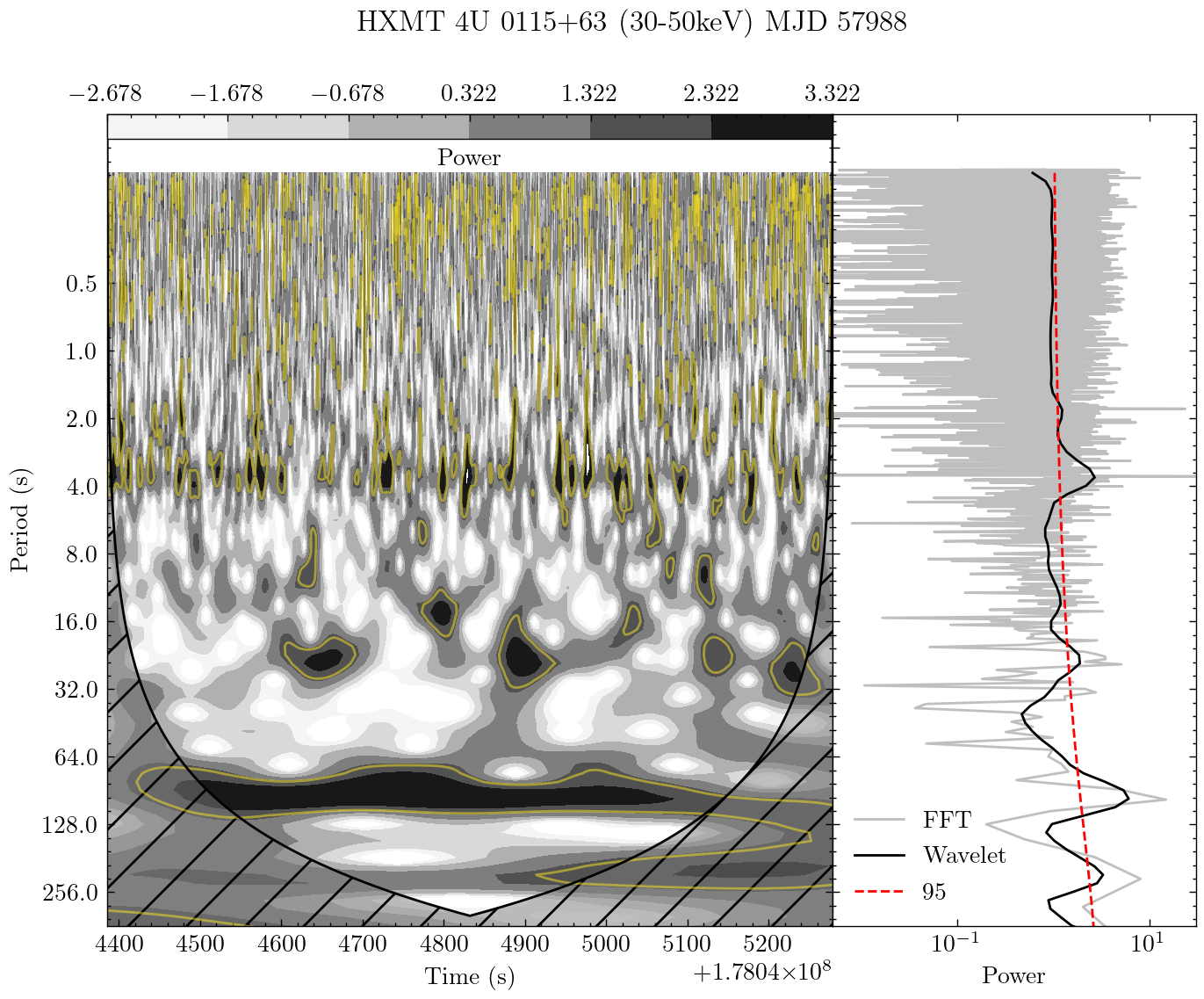}
		\end{minipage}
	}%
	\subfigure[]{
		\begin{minipage}[t]{0.5\linewidth}
			\centering
			\includegraphics[width=9cm]{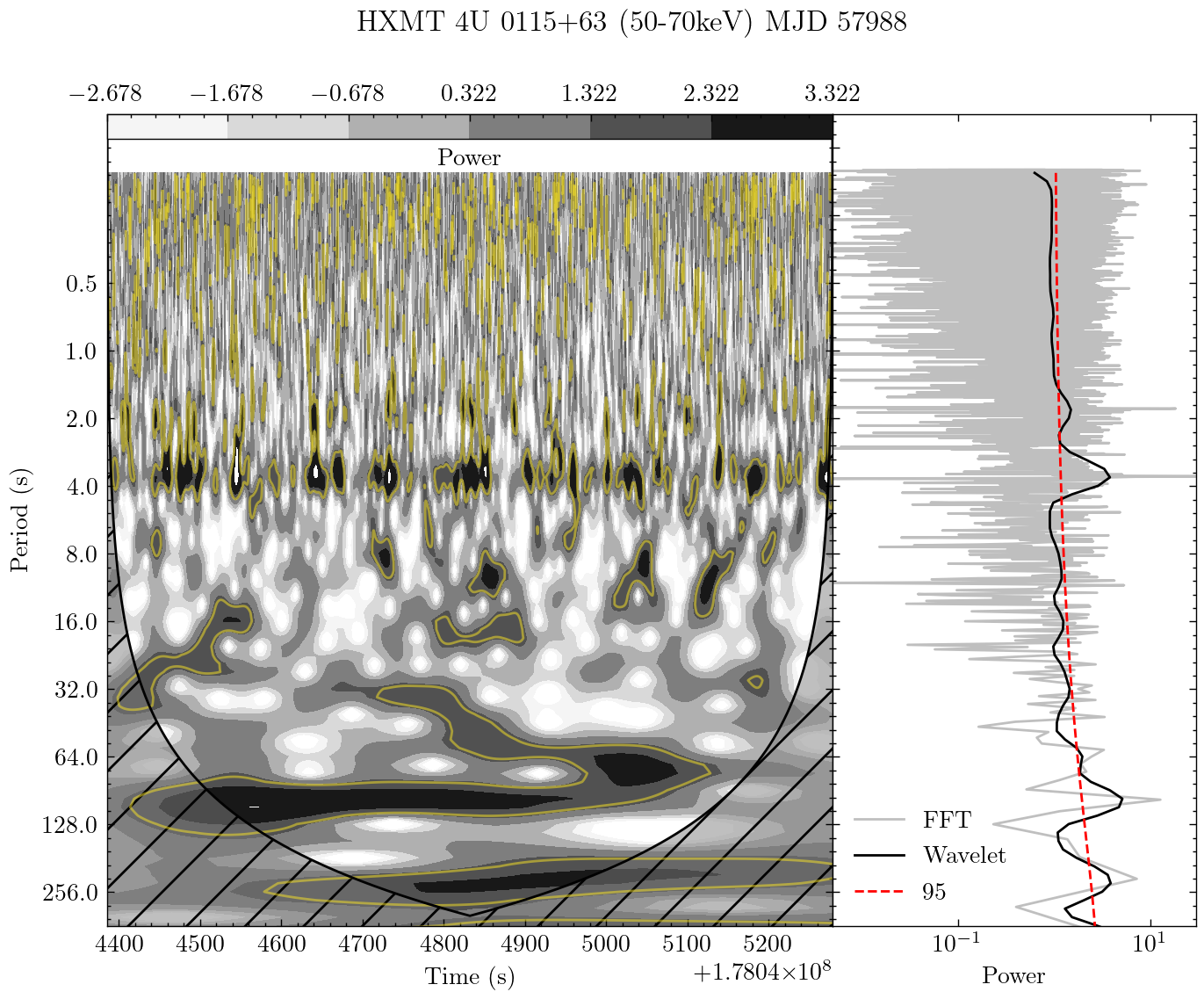}
		\end{minipage}
	}%

	\centering
	\caption{Sub-figures: Global wavelet spectra (right corners) and its contour plots (left corners) in MJD 57988 of 1-10 keV (a), 10-20 keV (b), 30-50 keV (c) and 50-70 keV (d). Yellow lines in contours refer to 95\% confidence spectrum; hashed area represents COI. Wavelet bases have been normalized in accordance with equation~\ref{eqn:norm} and their power has been transferred to log base two. Counts are normalized to speed up the transform. Data are segmented due to good time correction. (a) significant $ 100^{+20}_{-35}$ s QPO appeared (b) 100 s oscillation is still very prominent while neutron spin period can be seen, suggesting that radiation from accretion column may be predominant in this energy (10-20keV). More importantly, $ \sim $14.9 mHz (67 s), $ \sim $41 mHz (22 s) $\sim$62 mHz (16 s) QPOs are also detected simultaneously. (c) Most of the QPO features disappeared except for 100 s and 22 s oscillations. (d) 100 s QPO is still prominent and has the same profile in 50-70 keV while others may have been covered by noise. Almost no QPOs emerge in higher energy ($\gtrapprox 70$ keV). In contrast to the low luminosity, the $\sim100$ s QPO in this day had large Q-factor and S-factor.}	
	\label{fig:qpos}
\end{figure*}

\begin{figure*}
	\centering
	
	\subfigure[]{
		\begin{minipage}[t]{0.5\linewidth}
			\centering
			\includegraphics[width=8cm]{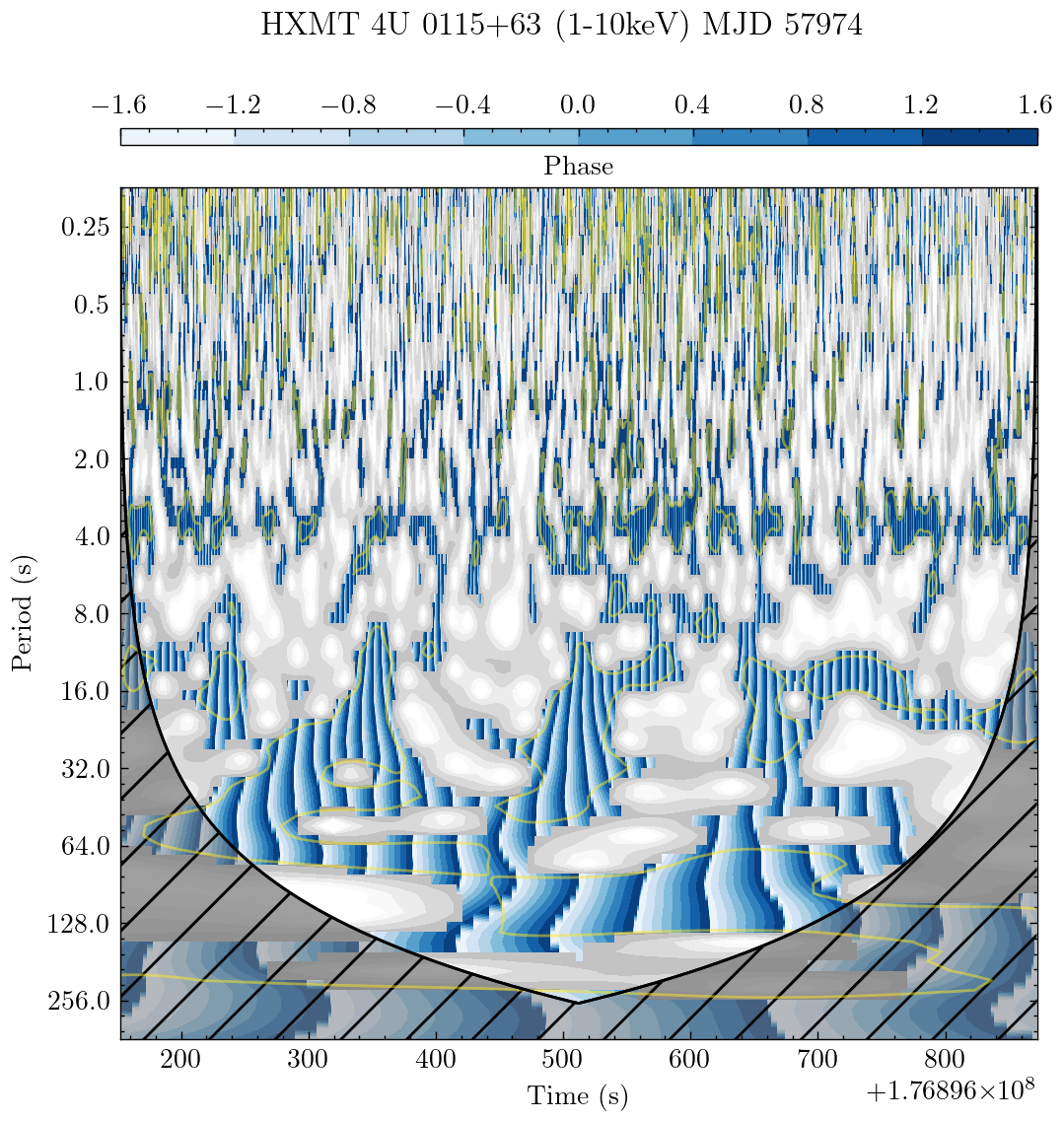}
		\end{minipage}%
	}%
	\subfigure[]{
		\begin{minipage}[t]{0.5\linewidth}
			\centering
			\includegraphics[width=8cm]{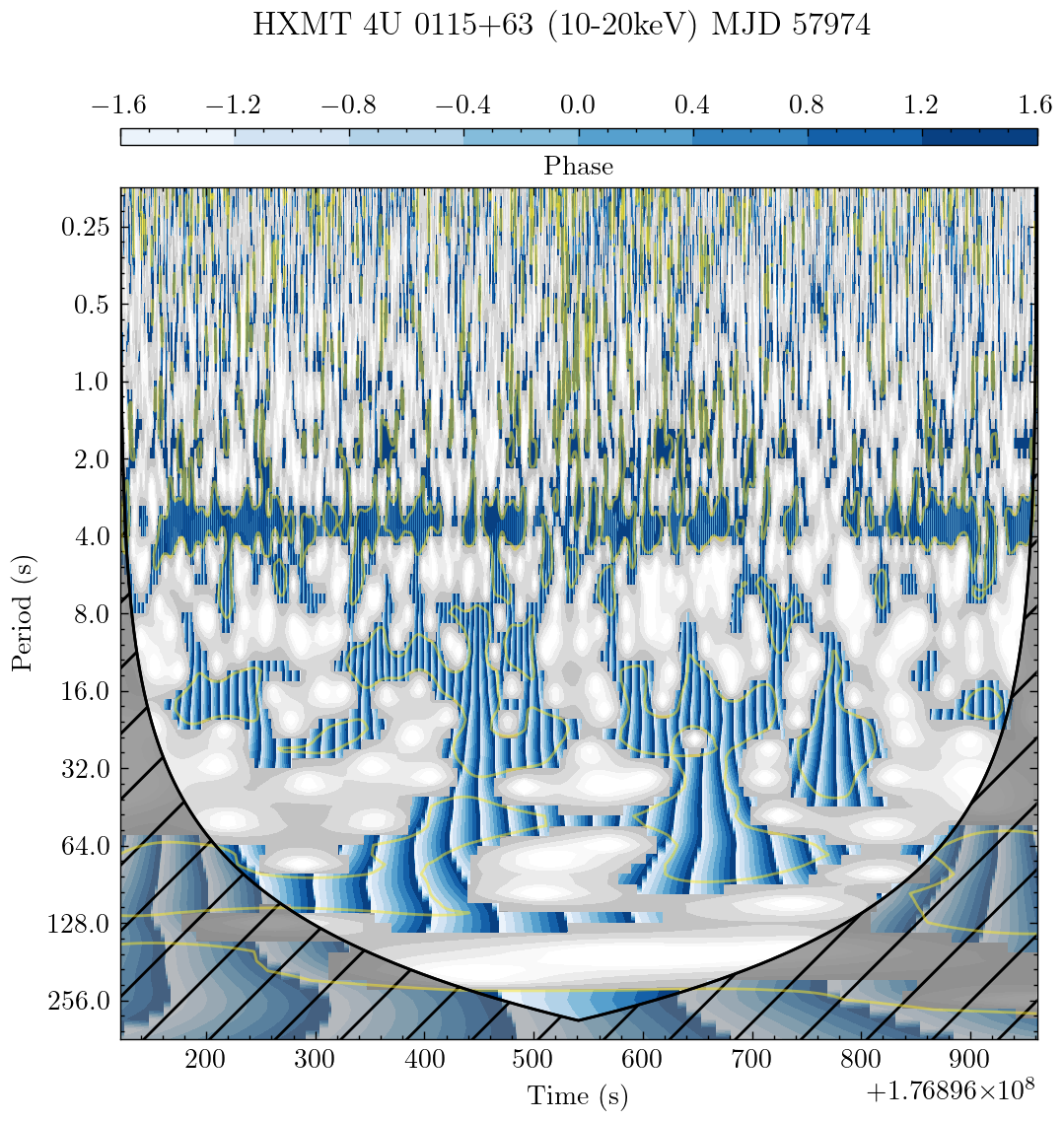}
		\end{minipage}%
	}%
	
	\subfigure[]{
		\begin{minipage}[t]{0.5\linewidth}
			\centering
			\includegraphics[width=8cm]{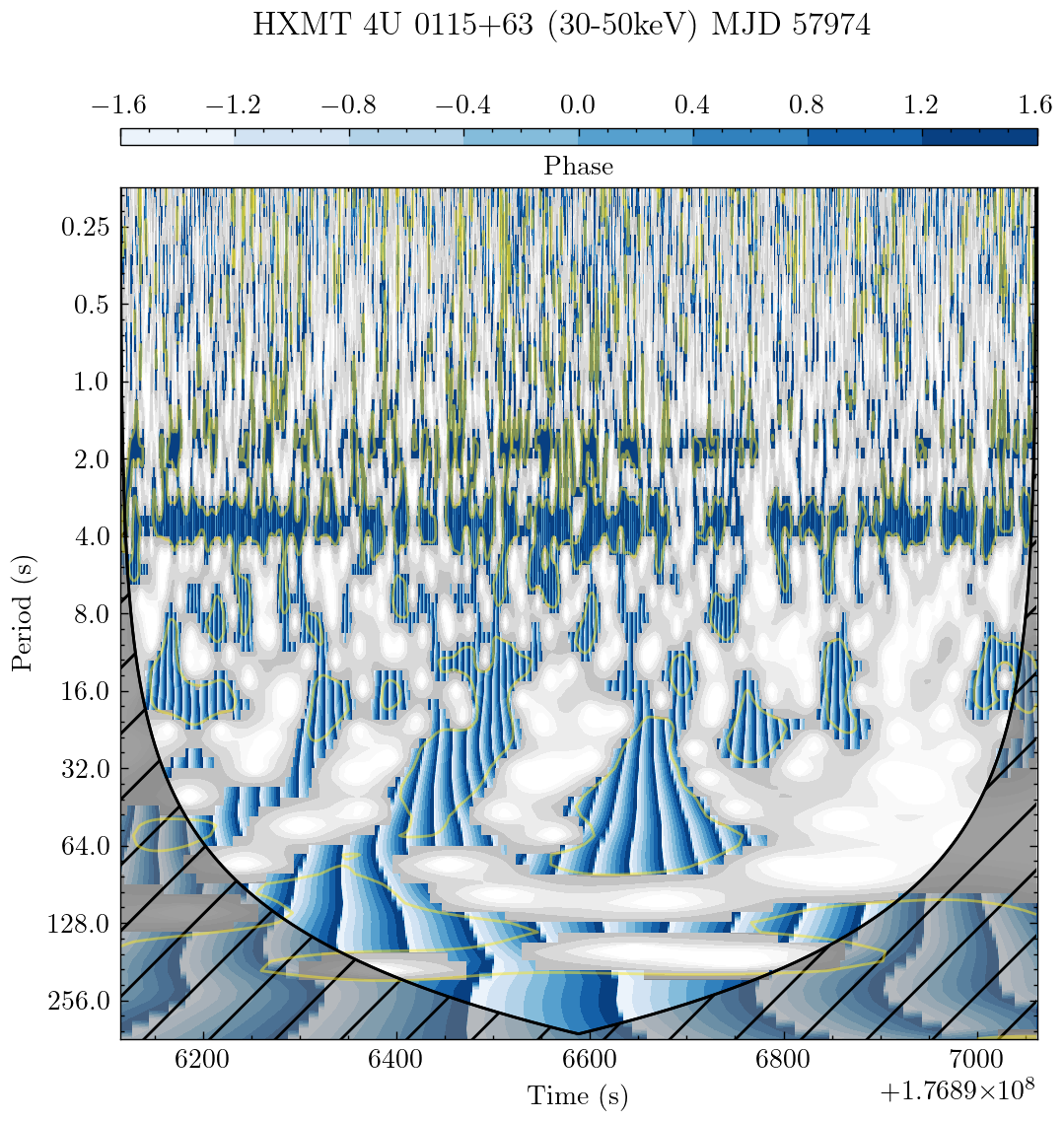}
		\end{minipage}
	}%
	\subfigure[]{
		\begin{minipage}[t]{0.5\linewidth}
			\centering
			\includegraphics[width=8cm]{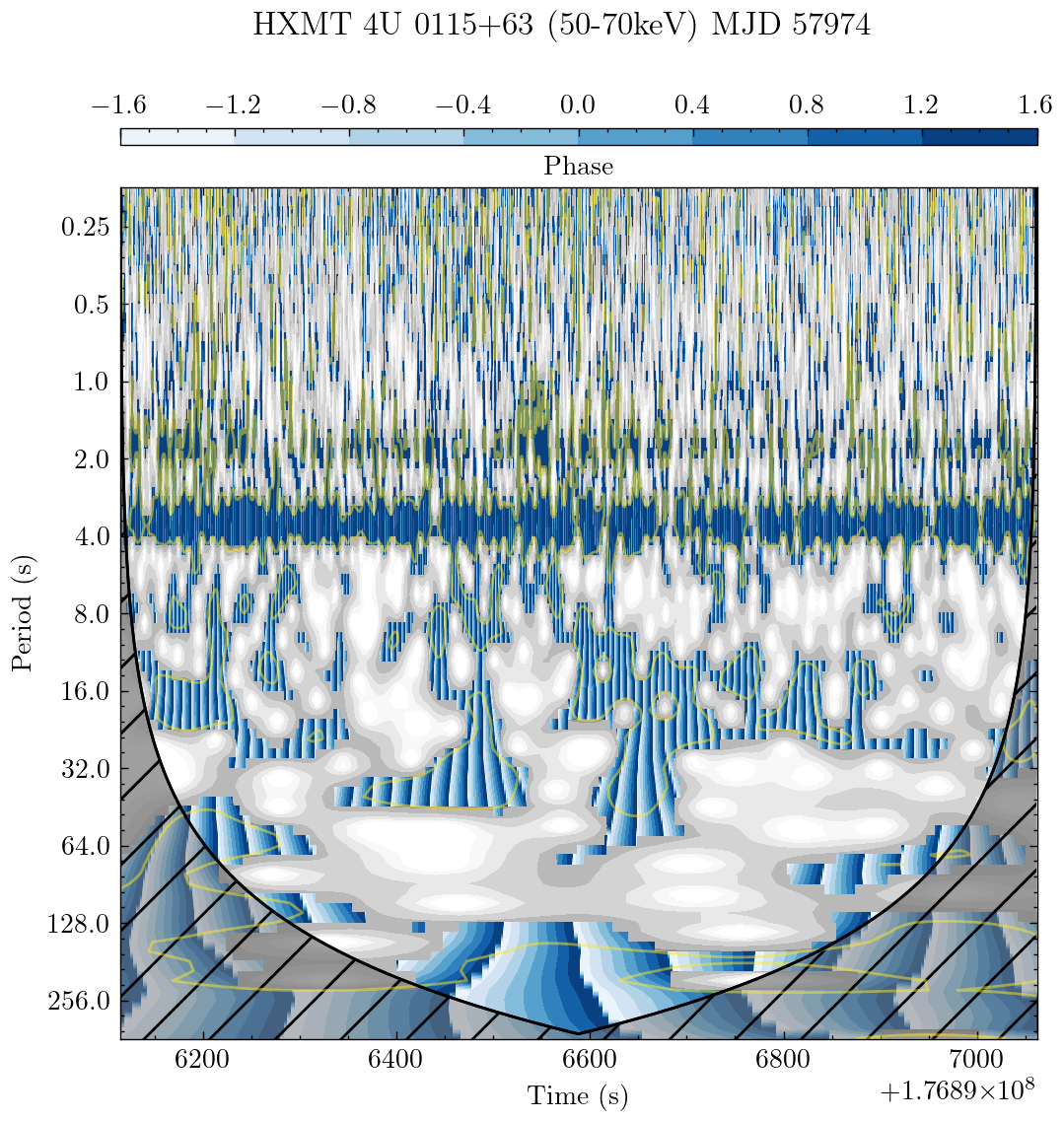}
		\end{minipage}
	}%
	
	\centering
	\caption{Phase maps of four different energy in MJD 57974 over-plotted on the wavelet power spectra. (a) 1-10 keV (b) 10-20 keV (c)30-50 keV (d) 50-70 keV. Original spectra are the same or akin to the left corner of the sub-figures of Figure~\ref{fig:qpos}. Blue areas with black lines showing the zero phase of pulsations. The observations of these maps are made at the same time (note X axes). It should be noted that the $\sim100$ s oscillation is translating from 50-70 keV to 1-10 keV over time(from (d) to (a)). Bi-directional pulsation phase drift from $\sim64$ s to $\sim40$ s and $\sim100$ s can be observed both in (a) and (b) about 400 s after the start of the exposure. Higher frequency QPOs (22-40 mHz) show almost no pulse phase or frequency drift, having a "jump-like" pattern.}
	\label{fig:qpophase}
\end{figure*}

Then by searching periods which attain half maximum power, it is also achievable to locate the main oscillation modes and their deviations. We include all these quantities in Table~\ref{tab:qpos}. Wavelet contours along with global spectra for MJD 57988 are plotted in Figure~\ref{fig:qpos}. To better understand the geometric properties of the sources of these oscillations, we made maps of the period phases which we over-plotted on the wavelet power contours. Examples of the most interesting maps have been presented in Figure~\ref{fig:qpophase}. These phases are calculated from daughter wavelets by:
\begin{equation}
	\phi=\arctan(\frac{Imag}{Real})
\end{equation}

\begin{figure}
	\includegraphics[width=\columnwidth]{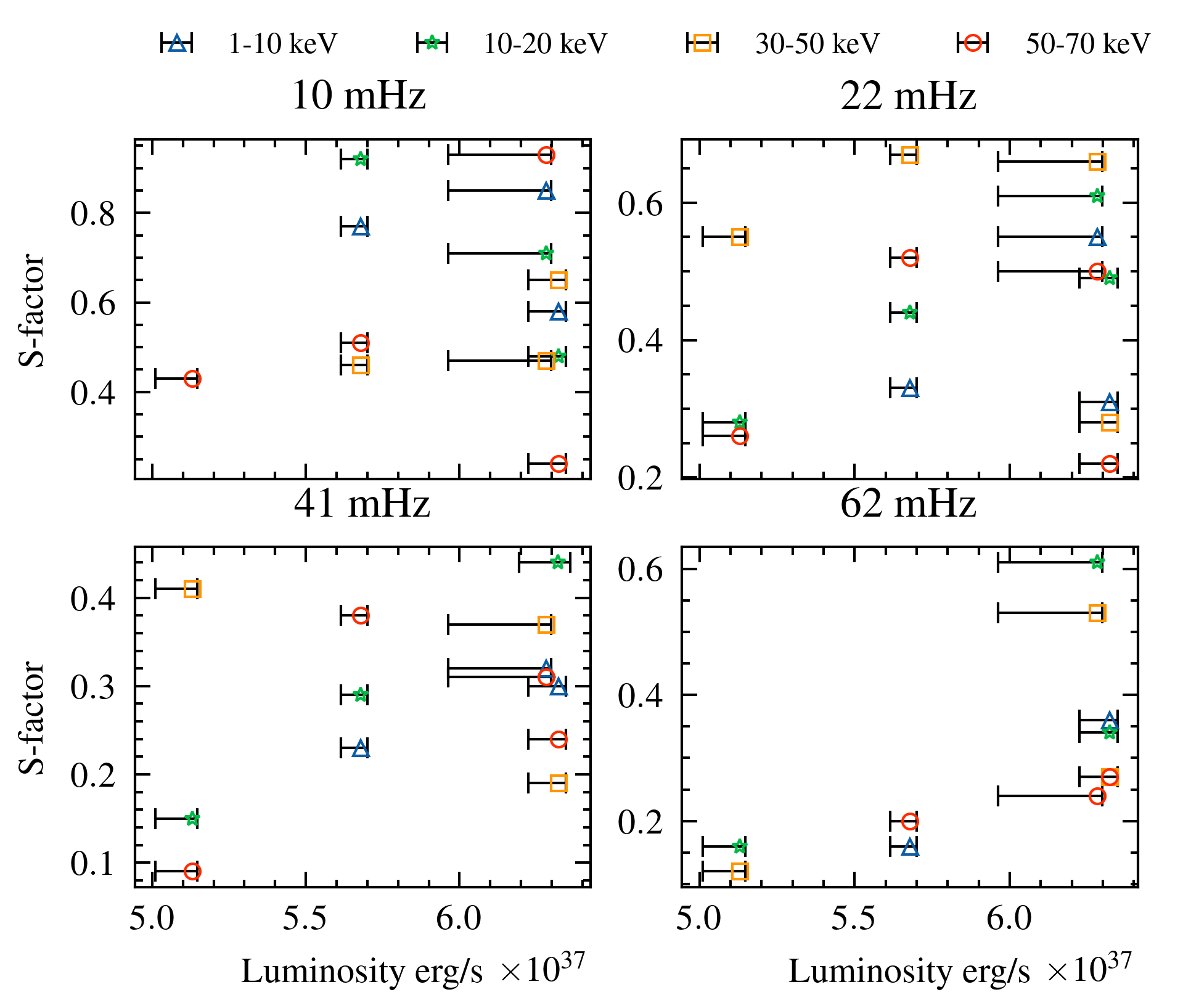}
	\caption{S-factor vs Luminosity. In general, most of these QPOs' stability is positive correlated with X-ray Luminosity, suggesting that the observational phenomenon is still correlated with the accretion rate. For lower frequency QPOs (i.e., 10 mHz, 22 mHz), the dropping of S-factor in brightest state is due to the shifting of oscillation with energy (see Figure~\ref{fig:qpophase}). Data in MJD 57988 was removed from to make the trend clearer.}
	\label{fig:lumivsstability}
\end{figure}

Results show a prominent QPO peak at $ \sim $10 mHz (100 s) during most of the observations (also see Figure~\ref{fig:qpos}). In addition, higher frequency QPOs at $ \sim $14.9 mHz (67 s), $ \sim $41 mHz (22 s) and $\sim$62 mHz (16 s) are also detected, which have been reported by \citet{dugair2013} and \citet{roy2019}. In wavelet spectrum contour (left corner), 22 s and 16 s QPOs are rather weak and highly scattered, having a stability factor lower than 0.5 and they emerged more at the brightest state while their stability dropped rapidly with the decreasing luminosity (see Figure~\ref{fig:lumivsstability}). However, their Q-factors show a significant negative correlation with luminosity.  It is also found that the $\sim100$ s QPO appeared in MJD 57988 illustrated a quite different feature compared with the same oscillation occurred in other outburst stage (see Figure~\ref{fig:qpos} and Table~\ref{tab:qpos}), having large Q-factor and S-factor. It should be noted that due to the transform edge effect, S-factor for low frequency QPOs (period $\geq 128$ s) may be underestimated, because padded zero would result in a reduced power inside COI. In the phase maps, one can found that QPOs with lower frequency ( $\lessapprox 22$ mHz) show phase drift frequently. In Figure~\ref{fig:qpophase}, for example, pulsation phase of $\sim64$ s oscillation drifted to $\sim40$ s and $\sim100$ s about 400 s after the start of exposure, which may imply a continuous shifting of the disc's geometric structure. A more interesting phenomenon was also observed in MJD 57974. In Figure~\ref{fig:qpophase}, the $\sim100$ s oscillation first occurred in 50-70 keV, then it disappeared and reappeared in 30-50 keV. At last, it prevailed in 1-10 keV. The oscillation was dissipating from 50-70 keV to 1-10 keV.

\begin{figure}
	\includegraphics[width=\columnwidth]{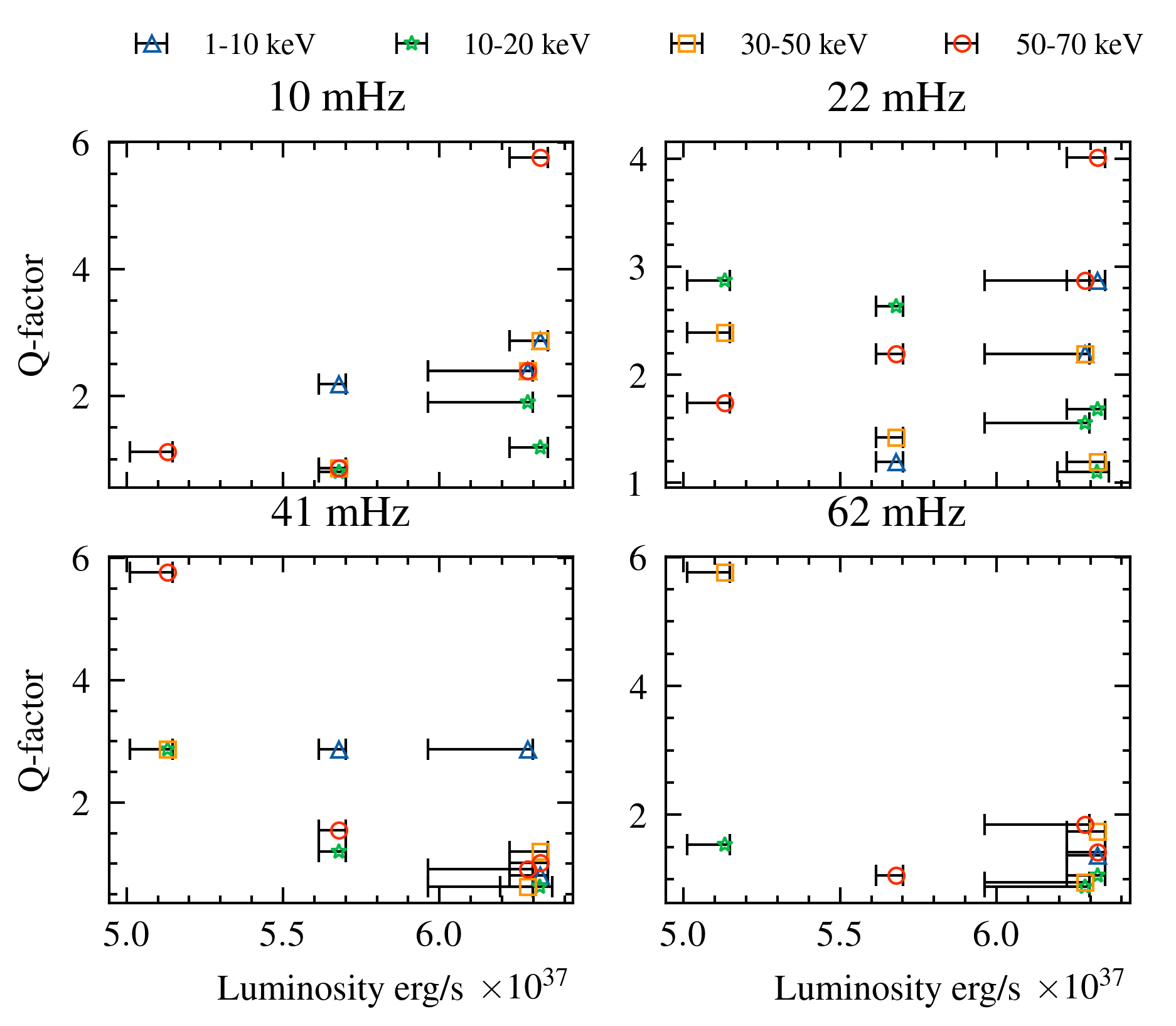}
	\caption{Q-factor vs Luminosity where QPOs having different frequencies for different energy bands illustrate a slightly different dependence on luminosity. Data in MJD 57988 was removed from this figure to make the trend clearer.}
	\label{fig:lumivsquality}
\end{figure}

Based on the calculated X-ray luminosity from 4-70 keV in \S 3.2 and the estimated Q-factors and S-factors in wavelet spectra, we plotted the relationships between Q/S-factors and X-ray luminosity (Figure~\ref{fig:lumivsquality} and Figure~\ref{fig:lumivsstability} ). Q-factors for the low frequencies (10 mHz and 22 mHz) showed a positive relation with X-ray luminosity, while in high frequency cases, it became a negative relation to luminosity for different energy bands. S-factors generally showed the positive relation with X-ray luminosity, while in highest luminosity range, it dropped due to the energy drifting of the QPO frequencies (see Figure~\ref{fig:qpophase} and Figures~\ref{fig:lumivsquality}, \ref{fig:lumivsstability}).

\begin{table*}
	\tiny
	\centering
	\caption{Summary of global parameters of QPOs observed in 4U 0115+63 by \textit{Insight}-HXMT during 2017 outburst. Q: quality factor, R: peak power relative to white noise spectrum, S: stability factor, total oscillation time at centroid period/frequency divided by exposure time. N refers to "Not detected". Deviations of centroid periods are estimated from FWHM. Very low frequency oscillations(Period $\gtrsim300$ s) are not detected from this analysis because these QPOs have dropped into the COI.}
	\label{tab:qpos}
	\renewcommand\arraystretch{2.2}
	\renewcommand\tabcolsep{3.0pt}
\begin{tabular}{cccccccccccccccccccccc}
	
	\multirow{2}{*}{\begin{tabular}[c]{@{}c@{}}Observation\\ (MJD)\end{tabular}} & \multirow{2}{*}{\begin{tabular}[c]{@{}c@{}}Energy\\(keV)\end{tabular}} &
	\multicolumn{4}{c}{200 s (4 mHz) QPO}	&\multicolumn{4}{c}{100 s (10 mHz) QPO} & \multicolumn{4}{c}{67 s (14.9 mHz) QPO} & \multicolumn{4}{c}{22 s (41 mHz) QPO} & \multicolumn{4}{c}{16 s (62 mHz) QPO} \\ \cline{3-22}
	&                                                                 & Period (s)       & Q&R       & S              & Period (s)     &  Q& R     & S      & Period (s)        & Q&R     & S       & Period (s)       &Q& R    & S     &Period (s)&Q&R&S \\ \hline
	\multirow{3}{*}{57967}
	& 10-20                                                                 &         $ 196^{+58}_{-69}$  &1.30&  11.80     &  0.89&N&N&N&N               & $ 53^{+10}_{-8} $   & 2.87&1.21  &  0.28  &       $ 22.5^{+4.2}_{-3.5} $        & 2.87 &  1.20 &0.15    &$10.3^{+5.6}_{-2.4} $   &1.54&1.30&0.16   \\
	& 30-50                                                                  &  $ 214^{+63}_{-86} $               & 1.10& 4.15     &  0.80  &N&N&N&N            &    $ 41^{+12}_{-6} $      & 2.39& 2.16    &   0.55     &    $ 20.7^{+3.9}_{-3.3} $        & 2.87&  1.80   &     0.41     &$ 10.33^{+0.93}_{-0.86} $  &5.76&1.00&0.12\\
	& 50-70                                                                &  $ 151^{+151}_{-24} $  &1.45&  4.03    &    0.76   &  $ 98^{+232}_{-15} $   &1.12&  2.29    &   0.43   &  $ 49^{+9}_{-14} $      &1.74&1.39        &   0.26   &$ 24.6^{+2.2}_{-2.0} $ &5.76&1.07&0.09 &N&N&N&N\\
	\hline
	\multirow{4}{*}{57968}
	& 1-10                                                                 &N&N&N&N          &       $ 139^{+26}_{-32} $      &2.19& 2.77     &   0.77     &        $ 75^{+14}_{-31} $  &1.19&1.94       &     0.33    &     $ 29.2^{+5.5}_{-4.7} $      &2.87 &1.21       &    0.23    &N&N&N&N\\
	& 10-20                                                                &N&N&N&N          &     $ 139^{+72}_{-41} $             &0.80&5.78     &   0.92     &          $ 53^{+4}_{-12} $         & 2.63& 2.10      &    0.44     &      $ 29^{+34}_{-7} $          &1.20 &  1.99      &  0.29       &N&N&N&N \\

	& 30-50                                                              &N&N&N&N            &     $ 90^{+17}_{-45} $     &0.86&2.57  &   0.46     &      $ 64^{+43}_{-15} $             & 1.42&3.38     &  0.67       &  N&N&N& N &N&N&N&N \\
	& 50-70                                                               &$ 180^{+34}_{-29} $ &2.87&4.43&0.78           &     $ 90^{+17}_{-45} $  &0.86    & 1.75    &   0.51     &         $64^{+12}_{-14} $      & 2.19&   2.49    &     0.52  &$ 29.2^{+8.7}_{-8.6} $ &1.55&2.13 &0.38 &$ 15^{+27}_{-3} $ &1.06&1.39&0.20 \\
	\hline
	\multirow{4}{*}{57971}
	& 1-10                                                                  &N&N&N&N         &  $ 127.45^{+24}_{-20} $&2.39& 5.59    &    0.85  &     $ 58^{+11}_{-13} $     &2.19 & 2.11      &   0.55      &    $ 22.5^{+4.3}_{-3.6} $   & 2.87 &1.40     &    0.32   &N&N&N &N  \\
	& 10-20                                                                    &$ 255^{+48}_{-58} $ &2.19&1.91&1.00     &           $ 107^{+32}_{-25} $      &1.90&    4.18     &   0.71     &  $ 53^{+16}_{-16} $      &1.55 & 2.53   &   0.61  &N&N&N&N    &$ 17^{+14}_{-7} $ &0.88&2.50&0.61  \\
	& 30-50                                                      &$ 255^{+48}_{-58} $ &2.19&2.04&1.00                    &    $ 98^{+29}_{-16} $  &2.39 &   3.14     & 0.47  &       $ 54^{+10}_{-12} $      &2.19 & 3.44    &   0.66   &  $ 27^{+43}_{-13} $ &0.62&2.49     &0.37     &$ 17^{+52}_{-4} $ &0.95&2.66&0.53 \\
	& 50-70                                                          &$ 255^{+48}_{-58} $ &2.19&6.28&1.00               &  $ 107^{+32}_{-17} $     &2.39  &4.61   &0.93        &  $ 53^{+10}_{-9} $    &2.87   &2.27   &0.50   &  $ 31^{+133}_{-7} $   &0.91 &1.69   &0.31  &$ 13.4^{+7.3}_{-2.1} $ &1.85&1.89&0.24      \\
	\hline
	\multirow{1}{*}{57972}
	& 10-20                                &$ 303^{+90}_{-69} $&1.90&3.14&1.00       &  N&N  &N &   N&    $ 76^{+23}_{-31} $ &1.10&  2.74  &0.38         &        $ 27^{+80}_{-12} $ &0.63      &  1.93  &   0.44     &N&N&N&N  \\
		\hline
	\multirow{4}{*}{57974}
& 1-10                               &N &N&N&N    &   $ 117^{+22}_{-19} $&2.87&1.84 & 0.58&    $ 45.1^{+8.5}_{-7.2} $ &2.87&1.43 &0.31         &        $ 25^{+7}_{-12} $       &0.81&1.72  &0.30    & $ 14^{+17}_{-2} $ &1.37&1.76&0.36  \\
& 10-20                         &N&N&N&N  &$ 98^{+19}_{-40} $&1.19 &1.71 &0.48     & $ 69^{+47}_{-11} $  &1.68&1.64 &0.49  &    N&N&N&N  &$ 19^{+13}_{-6} $ &1.06&1.84&0.34  \\
& 30-50                                         &N&N&N&N        &    $ 139^{+26}_{-22} $ &2.87& 2.00 & 0.65   &   $ 58^{+11}_{-24} $   &1.19& 1.47  &  0.28       & $ 27^{+49}_{-4} $  & 1.20&1.33  &0.19  &$ 17.4^{+3.3}_{-5.1} $ &1.74&1.57&0.27  \\
& 50-70                                     &N&N&N&N     &$ 127^{+12}_{-11} $&5.76&1.08&0.24             &$ 37.9^{+7.2}_{-3.1} $&4.01&1.21&0.22    &$ 23^{+4}_{-10} $&1.01&1.46&0.24   &$16 ^{+11}_{-4} $ &1.42&1.58&0.27   \\
\hline
	\multirow{4}{*}{57977}
	& 1-10                               &N &N&N&N    & $ 128^{+24}_{-20} $  &2.87&1.91 & 0.74&    N &N& N &N         &        $ 29.2^{+5.5}_{-8.6} $       & 1.74& 1.25  &   0.27    &N &N&N&N  \\
	& 10-20                &         $ 165^{+31}_{-26} $&2.87        &4.57         &  0.89     & $ 76^{+31}_{-38} $  &0.77 &  1.66   &0.46    &    $49^{+49}_{-8} $   &1.45 &    2.09     & 0.47   &$ 29^{+9}_{-12} $ &1.10&1.95&0.34   &$ 21^{+17}_{-5} $ &1.33&1.89&0.36  \\
	& 30-50                                         &N&N&N&N        &    $ 139^{+26}_{-32} $ & 2.19& 5.78    &   0.91   &   $ 38^{+7}_{-19} $   &0.86& 2.16       &  0.42       & $ 25^{+25}_{-7} $  & 1.09& 1.84  &0.29  &$ 11.3^{+3.3}_{-1.8} $ &2.39&1.45&0.18  \\
	& 50-70                                     &   $ 152^{+29}_{-82} $   &0.75&  2.08   & 0.67    &   $ 98^{+67}_{-23} $      &1.42 &   3.14      &  0.63 &N&N&N&N               &   $ 29^{+16}_{-10} $   &1.12&  1.67    & 0.24   &$ 12.3^{+2.3}_{-2.0} $ &2.87&1.27&0.18   \\
		\hline
	\multirow{4}{*}{57988}
	& 1-10                                                           &N&N&N&N                &   $ 90^{+17}_{-14} $&2.87& 2.66 &  0.70     &     $ 49.1^{+9.3}_{-4.1} $       & 4.01 &1.12    &0.22    &     N  &N& N   &   N &N&N&N&N  \\
	& 10-20                                &$ 303^{+57}_{-69} $&2.19&1.72&0.76       &  $ 98^{+19}_{-23} $  &2.19&4.60 &0.85&    $ 45^{+9}_{-21} $ &1.01&1.58  &0.40         &        $ 29^{+24}_{-5} $     &1.55 &  1.56  &   0.24     &$ 13.4^{+2.5}_{-2.1} $&2.87&1.21&0.13  \\
	& 30-50                                         &$ 214^{+19}_{-18} $ &5.76&1.41&0.41        &   $ 98^{+19}_{-16} $   &2.87 &   3.18 &0.82    & N    & N  &N   &N   & $ 25^{+2.2}_{-3.9} $         &3.67&  1.31  &  0.24   &N&N&N&N  \\
	& 50-70                                      &$ 234^{+21}_{-37} $ &3.67&1.60&0.72        &  $ 98^{+19}_{-16} $ &2.87  & 2.64   &0.71&$ 63.7^{+5.8}_{-5.3} $   &5.76& 1.12& 0.27 &  N& N&  N & N&N&N&N&N   \\
	\hline
\end{tabular}
\end{table*}


\section{discussion}
In the previous section, we reported the detection of a new low-frequency QPO feature in the HMXB pulsar 4U 0115+63 at $P_{\rm QPO}\approx 100 {\rm s}\ (\nu_{\rm QPO}\approx10 $ mHz). Narrow peaks corresponding to the spin period of the accretion-powered pulsar are clearly seen, both in the local and global wavelet spectrum. We also found the other QPO frequencies at $\sim$ 4 mHz, 15 mHz and 41 mHz and 62 mHz in the multi-band light curves of 4U 0115+63. Low-frequency QPOs of $\sim 10-100$ mHZ occur commonly in HMXBs. Here, we aim to present possible theoretical explanation for the QPOs discovered in the previous section.

There have been various models proposed to explain QPOs features in HMXBs. Among them, the most popular ones are Keplerian Frequency Mode (KFM) \citep{vanderklis1987} and Beat Frequency Model (BFM) \citep{alpar1985}. In KFM, QPOs result from the inhomogeneities in the inner accretion disc  at the Keplerian frequency while QPOs in BFM is the difference between the spin frequency and the Keplerian frequency of the inner edge of the accretion disc, meaning that $\nu_{\rm QPO}=\nu_k-\nu_s$. The accretion flow inside the Alfv\'en radius is expected to flow along the field line, so one may expect the inner edge of accretion disc corresponding to its magnetospheric radius \citep{ghosh1979,becker2012}.
\begin{equation}
	\begin{aligned}
		R_m=273(\frac{\Lambda}{0.1})(\frac{M_{\star}}{1.4M_{\sun}})^{1/7}(\frac{R_\star}{10 \,\rm km})^{10/7}\\
		\times (\frac{B_\star}{10^{12}G})^{4/7}(\frac{L_x}{10^{37}\rm\,erg\,s^{-1}})^{-2/7}\;\rm km,
		\label{eqn:rm}
	\end{aligned}
\end{equation}
where $M_\star$ and $R_\star$ refer to the mass and radius of the neutron star and $L_x$ to the X-ray luminosity. $\Lambda$ is a constant and is 1 for spherical accretion. For disc accretion, it can be approximated as $\Lambda\approx0.22\alpha^{18/69}$ \citep{harding1984,becker2012}. Keplerian frequency in this radius can be calculated immediately:
\begin{equation}
	\nu_k=\frac{1}{2\pi}\sqrt{GM/R^3}\approx15\, \rm Hz,
\end{equation}
As the pulsar spin frequency is given as $\nu_s \approx 1/3.61=0.277$ Hz, it is then apparent that the observed periods are much smaller than the oscillation period for both KFM and BFM in 4U 0115+63.
 \citet{roy2019} used thermal disc instabilities model to explain 1 mHz and 2 mHz oscillations while \citet{angelo2010} using 1-D simulation showed the oscillation frequency of the inner accretion disc radius spanning three orders of magnitude, from 0.02 to 20$t_{\rm visc}$. A viscosity timescale derived from classical $\alpha$-disc model has been used in their work:
\begin{equation}
t_{\rm visc}=991.1L_{37}^{-23/35}\alpha_{0.1}^{-109/230}R_{6}^{52/35}m_{1.4}^{-51/70}\,\rm s
\label{viscqpo}
\end{equation}
It should be noted that $L_{37}=\frac{L_{x}}{10^{37}\;\rm erg/s}$ and $R_{6}=\frac{R_{m}}{10^6\;\rm cm}$, $ m_{1.4}=\frac{M_{\star}}{1.4M_{\sun}} $ while $\alpha_{0.1}=\frac{\alpha}{0.1}$ is the classical $\alpha$  viscosity parameter. Their theory explained low frequency oscillations in 1-2 mHz well. In addition, considering the possibility suggested by \citet{shirakawa2002}, they proposed that a warped or precessing accretion disc caused by misaligned vectors of the disc angular momentum and the magnetic field can also lead to such phenomenon. The frequency of warping/precession at the magnetospheric radius lies in the milihertz range and the corresponding precession timescale are given in their paper:
\begin{equation}
t_{\rm prec}\approx775.9(\frac{L_x}{10^{37}\rm \,erg/s   }   )   ^{-0.71}(\frac{\alpha}{0.1})^{0.85}\;\rm s.
\label{eqn:magqpo}
\end{equation}
These models predict a positive correlation between QPO centroid frequency and the X-ray luminosity and they fit both the $\sim$ 1 and $\sim$ 2 mHz oscillations well. However, this correlation was not seen in higher frequency QPOs. It may imply that the features are not caused by these mechanisms. S-factors of these QPOs show a positive dependence on the X-ray luminosity, while there are no apparent correlations between their properties and X-ray hardness. Q-factor for high frequency oscillations ($P\lessapprox 50$ s) and that for lower frequency have slightly different dependence on accretion rate. It should be noted that if the $\sim 100$ s oscillation is due to the inhomogeneity, the oscillation should be generated approximately in
\begin{equation}
	R_{\rm QPO}=(\frac{GM_\star}{4\pi^2(\nu_{\rm QPO}+\nu_s)})^{1/3}
	\label{eqn:Rqpo}
\end{equation}
The radius obtained for a mass of $1.4M_{\sun}$  is $2.6\times10^8 $ cm, an order of magnitude larger than magnetospheric radius. A more serious problem resides in the coexistence of QPOs with different frequencies. It may imply that we cannot keep our indifference of the fine structure of inner edge anymore.  Describing physical processes occurred inside the neutron star magnetospheric radius has long been a challenging issue, but with the observational phenomena we have revealed with wavelet transform, a hope of solving this intriguing problem may arise. Almost all previous magnetic hydrodynamic (MHD) simulation assumed the configuration of inner disc is axisymmetric, but observations suggest that there may be periodic structure on the inner disc. Moreover, there is indeed a mechanism in hydrodynamics that can induce periodic structure in swirling flows. We now present an interesting possibility to understand these sub-Hz oscillations.

First, one knows that the magnetic field pattern rotate with an angular velocity approximately equals to the same figure of neutron star, $\omega_\star$. So if the Kepler angular velocity at $R_m$ slightly lower than $\omega_\star$, particles attached to field lines would spiral outwards to larger \textit{R}, repelled by the centrifugal force. we can put this requirement in another way by referring to the corotation radius \citep{frank2002}.
\begin{equation}
	R_C=(\frac{GMP^2_{\rm spin}}{4\pi^2})^{1/3}=1.5\times10^8P^{2/3}_{\rm spin}m_1^{1/3}\rm \,cm,
\end{equation}
where $P_{\rm spin}$ refers to the spin period of central neutron star. A steady accretion flow required $R_{\rm C}>R_{\rm M}$.  Equating the magnetic pressure to the ram pressure from spherically symmetric gas in free-fall on to the star, one could obtain the Alfv\'en radius.
\begin{equation}
	r_M=2.9\times10^8m_1^{1/7}R_6^{-2/7}L_{37}^{-2/7}\mu_{30}^{4/7}\,\rm cm
\end{equation}
where $m_1$ is the mass of neutron star in terms of solar mass, $\mu_{30}$ is dipole moment measured in $10^{30}\,G\,\rm cm^{3}$.
Several estimates of $R_m$ have been made but all give similar answers, of order the spherical Alfv\'en radius \citep{frank2002,angelo2010}:
\begin{equation}
	R_M\sim0.5r_M
\end{equation}
Equal this radius with $R_C$, the equilibrium spin period $P_{\rm eq}$ can be found and it is a highly suggestive result:
\begin{equation}
	P_{\rm eq}\approx3m_1^{-2/7}R_6^{-3/7}L_{37}^{-3/7}\mu_{30}^{6/7}\,\rm s
\end{equation}
This approximately equals to the spin period of neutron star in 4U 0115+63. So the disc is likely to be truncated at the corotation radius and the accretion rate would be greatly affected by neutron star's magnetosphere. Outside the corotation radius or magnetospheric radius but inside the Alfv\'en radius, angular momentum was transformed into the disc material by the strong magnetic field, accelerating them considerably. As \citet{angelo2010} has shown, we can expect the inner region of the disc being pushed outwards by centrifugal force and restrained by a dense material boundary at a larger radius. An interface of very interesting properties was born there.

One should be aware that swirling flows in this condition are susceptible to instabilities that often lead to the roll up of vorticity filament into vortices via Kelvin-Helmholtz instability. Since we have a concrete boundary outside the thin material layer and a centrifugal force field, the  interface is likely to have a meta-stable polygon-shape configuration having vortices at each vertice, a phenomenon of symmetry breaking which have been confirmed in laboratories \citep[see][]{kurakin2002,vatistas2008,abderrahmane2013}. In 4U 0115+63, magnetic field acts as the rotating plate. At different phases of the outburst, varying viscosity ($\alpha$-parameter), radiation pressure, disc material density and velocity are likely to induce very different geometric configurations and the regular periodic structure is possible to be destructing and reconstructing alternatively, which can account for the transient nature of these sub-Hz QPOs. The drifting of $\sim100$ s oscillations could be explained by the shifting between different polygon modes. Moreover, $\sim41-62$ mHz ($P_{\rm QPO}\sim10-30$ s) oscillations may come from sub-vortices at the inner edge. Since we observe almost no frequency drift of these QPOs, vortices here is expected to behave like the defections in a bearing, which can account for the "jump-like" pattern they showed in wavelet power spectra. If the structure of the inner interface do take the expected configuration, it may be able to estimate the rotation speed, number of sub-vortices and radius of the inner edge through the characteristics of the sub-Hz QPOs. This model can also explain qualitatively the anomaly stable $\sim100$ s oscillation in MJD 57988, when the luminosity or accretion rate dropped to low levels.  Although there are all the essential ingredients to produce this polygon-like pattern, the multi-vortex model is a hypothesis at present. Precise mathematical derivation is needed and detailed hydrodynamic simulation should be conducted to confirm it.

\section{Conclusions}

Neutron stars in Be X-ray binaries act as a productive space laboratory, providing a closer glance into the physics in extreme environment. Its outburst conveyed a bunch of valuable information about the physics process inside the inner edge of accretion disc. We observed the outburst of 4U 0115+63 during 2017 August using  \textit{Insight}-HXMT, performed both timing and spectral analysis. We confirmed that the source is still spinning up at a moderate rate during the outburst and the apsidal motion is also confirmed to be $\dot{\rm\omega}=0^{\circ}.048\pm0^{\circ}.003\; \rm yr^{-1}$. The evolution of pulse profile has been illustrated in details.

We reported a newly discovered $\sim100$ s (10 mHz) QPO during 2017 outburst and also confirmed multiple QPOs that have been observed previously. Wavelet analysis is introduced in our QPO studies, and we defined three characteristic quantities: Q-factor, R-factor, and S-factor to facilitate our analysis. With the help of its locality, one is able to investigate the stability and distribution of the abundant sub-Hz QPO features in 4U 0115+63. While centroid period/frequency has almost no correlation with luminosity, the S-factors of these sub-Hz oscillations are likely to be positively correlated with accretion rate. In addition, it is possible that in brightest state the system became more noisy in high frequency, as the Q-factors of high frequency QPOs depend on the luminosity negatively and S-factors show a opposite dependence. For $\sim 10$ mHz QPO, these characteristic factors show a weak positive dependence on source luminosity, but are abnormally high in MJD 57988. In the brightest state, MJD 57974, the $\sim100$ s oscillation translated from 50-70 keV to 1-10 keV over time, resulting a  sharp drop of both S-factor and Q-factor. On grounds of the phase maps and wavelet spectrum, we conduct a brief and clear reasoning to show the possible generating mechanism of these QPOs. The multi-vortices model is promising and is possible to account for the complex QPOs features in 4U 0115+63. Then, it is of the utmost importance to conduct a hydrodynamic simulation at the inner edge to check this picture.

\section*{Acknowledgements}

This work is supported by the National Natural Science Foundation of China (Grants No. U1838103, 11622326, U1838201, U1838202), the National Program on Key Research and Development Project (Grants No. 2016YFA0400803, 2016YFA0400800) and the College Students' Innovative Entrepreneurial Training Plan Program of China (project number:S202010486293). This work made use of data from the \textit{Insight}-HXMT mission, a project funded by China National Space Administration (CNSA) and the Chinese Academy of Sciences (CAS).

\section*{Data Availability}

Data that were used in this paper are from Institute of High Energy Physics Chinese Academy of Sciences (IHEPCAS) and are publicly available for download from the \textit{Insight}-HXMT website.
To process and fit the spectrum and obtain folded light curves, this research has made use of HEASoft (XRONOS and FTOOLS) provided by NASA.



\bibliographystyle{mnras}
\bibliography{example} 




\appendix
\section{Wavelet analysis procedure}
\label{sec:wavelet}
We have chosen Morlet as our wavelet base to perform further analysis. "Wavelet basis" refers to a set of orthogonal functions, which is often used in continuous wavelet transform. In this paper, only the continuous transform is used.The continuous wavelet transform of a discrete sequence $x_n$ is defined as the convolution of $ x_n $ with a scaled and translated version of bases $ \psi_0(\eta) $
\begin{equation}
W_n(s)=\sum_{n'=0}^{N-1}x_{n'}\psi^{*}[\frac{(n'-n)\delta t}{s}],
\label{eqn:trans1}
\end{equation}

The subscript "0" has been dropped to indicate that these bases have been normalized. * refers to the complex conjugate. Daughter wavelets are varied with the wavelet scale \textit{s} and translated along the localized time index \textit{n}. $\delta$t refers to sampling rate. Morlet wavelet can be expressed as:

\begin{equation}
\psi_0(\eta)=\pi^{-1/4}e^{i\omega_0\eta}e^{-\eta^2/2},
\end{equation}

$\omega_0$ is the non-dimensional frequency, being taken to be 6 to satisfy the admissibility condition \citep{farge1992}. To realize the continuous wavelet transform, if the number of points in times series is denoted by N, the discrete convolution should be done N times for each scale. Applying a Discrete Fourier transform(DFT) or Fast Fourier transform(FFT), time series in fourier space can be denoted by $\hat{x}_k$. $k=0...N-1$ is the frequency index. Taking the continuous limit, the Fourier transform of a function $\psi(t/s)$ is given by $\hat{\psi}(s\omega)$. Besides, by the convolution theorem, the wavelet transform is just the inverse Fourier transform of the product:

\begin{equation}
W_n(s)=\sum_{k}^{N-1}\hat{x}_k\hat{\psi}^{*}(s\omega_k)e^{i\omega_kn\delta t},
\label{eqn:wavelettransform}
\end{equation}

where the angular frequency is defined as:

\begin{equation}
\omega_k=
\begin{cases}
&\frac{2\pi k}{N\delta t}:k\le\frac{N}{2}\\
-&\frac{2\pi k}{N\delta t}:k>\frac{N}{2}
\end{cases}
\label{eqn:transfrequency}
\end{equation}

As is mentioned previously, all bases have been normalized to ensure that the wavelet transforms at each scale are comparable with each other. Specifically, the wavelet functions at each scale have been normalized to have unit energy, as given by TC98:

\begin{equation}
\hat{\psi}(s\omega_k)=(\frac{2\pi s}{\delta t})^{1/2}\hat{\psi}_0(s\omega_k)
\label{eqn:norm}
\end{equation}
Then, at each scale \textit{s}, one should have:
\begin{equation}
\sum_{k=0}^{N-1}|\hat{\psi}(s\omega_k)|^2=N
\end{equation}

As a consequence, the wavelet transform will be weighted only by the amplitude of the Fourier coefficients $\hat{x}_k$. The wavelet power spectrum is defined as $|W_n(s)|^2$. Using the normalization mentioned in equation~\ref{eqn:norm}, the expectation value for $|W_n(s)|^2$ should be N times the expectation value for $|\hat{x}_k|^2$. For a white-noise time series, this expectation value is $\sigma^2/N$, where $\sigma^2$ is the variance, suggesting that the expectation value for its wavelet transform is $|W_n(s)|^2$=N. In this study, we assume white-noise only, thus one can compare the background spectrum with the result of wavelet transform. If a peak in the wavelet power spectrum is significantly above this background spectrum, then we could judge the confidence level of this signal to be a true feature. One should also notice that if noise $x_n$ is a normally distributed random variable, then its Fourier power spectrum should also normally distributed. So $|\hat{x}_k|^2$ is chi-square distributed with two degrees of freedom (DOFs) \citep{jenkins}. We then determine the 95\% confidence level.

It is essential to select a set of scales $s_j$ to use in our wavelet analysis. One is limited to chose a discrete set of scales in orthogonal transform \citep{farge1992}. As given by TC98, we have:
\begin{equation}
s_{j}=s_{0}2^{j\delta j},j=0,\,1,...,\,J
\end{equation}
Here, J is given by:
\begin{equation}
J=\delta j^{-1}log_2(N\delta t/s_0),
\end{equation}
$s_0$ here is the smallest resolvable scale and should be chosen so that the equivalent Fourier period is approximately $2\delta t$ (TC98). \textit{J} determines the largest scale, mainly influenced by the length of time series N. $\delta j$ controls the resolution of wavelet analysis. For the Morlet wavelet, a $\delta j$ of about 0.5 is the largest value that still provides adequate sampling in scale. In general, smaller $\delta j$ gives finer resolution. Here, we scaled as: 1)$s_0=2\delta t$; 2)$\delta j=0.05$; 3)$\delta t=0.078125s$.

Given that one is handling with finite-length time series, errors will appear inevitably. In this study, data are assumed to be cyclic so one should pad the end of the time series with zeroes before doing the wavelet transform so as to bring the total length N up to the next higher power of two, by which the edge effects are limited and Fourier transform speed can be sped up effectively. These zeros cause discontinuities at the end points and as one goes to larger scales (periods), amplitudes near the edges will decrease as more zeroes enter the analysis. The \textit{cone of influence} (COI) is a region of the wavelet spectrum in which edge effects become important. It is indicated in Figure~\ref{fig:qpos} and Figure~\ref{fig:qpophase} by the cross-hatched regions. The peaks within these regions have presumably been reduced in magnitude due to the zero padding. A 95\% confidence spectrum, constructed from a white noise background is used to distinguish fake signals. One can compare it with the result of wavelet transform.


\bsp	
\label{lastpage}
\end{document}